\documentclass{article}

\usepackage{natbib}
\usepackage{amsmath}
\usepackage{graphicx}
\usepackage{hyperref}{}
\usepackage{float}
\usepackage{txfonts}
\usepackage{amssymb}
\usepackage{mathrsfs}
\usepackage{stmaryrd}
\usepackage{multirow}
\usepackage{multicol}
\usepackage{subcaption}
\usepackage[labelfont=bf]{caption}
\usepackage{lmodern}
\usepackage{wrapfig}
\usepackage{blindtext}
\usepackage{enumitem}
\usepackage{amssymb}
\usepackage{esvect}
\usepackage{empheq}
\usepackage{cancel}
\usepackage{hhline}
\usepackage{natbib}
\usepackage{times}
\usepackage{color}
\usepackage{authblk}
\usepackage[margin=0.8in]{geometry}
\usepackage{fancyhdr}
\hypersetup{
    colorlinks=true,
    citecolor=blue,
    linkcolor=red,
    urlcolor=black
    }

\def\mearth{\,{\rm M}_\oplus}

\def\zgal{Z_{\rm Gal}}
\def\xgal{X_{\rm Gal}}
\def\ygal{Y_{\rm Gal}}
\def\zext{Z_{\rm ext}}

\def\ymean{\bar Y}

\begin{document}

\title{New models of Jupiter in the context of Juno and Galileo}

\renewcommand{\thefootnote}{\fnsymbol{footnote}}
\author[1,2]{Florian Debras \thanks{Corresponding author : \href{mailto:florian_debras@hotmail.com}{florian\_debras@hotmail.com} }}
\affil[1]{Ecole normale sup\'erieure de Lyon, CRAL, UMR CNRS 5574, 69364 Lyon Cedex 07,  France}
\affil[2]{School of Physics, University of Exeter, Exeter, EX4 4QL, UK}

\author[1,2]{Gilles Chabrier}

\maketitle

Accepted in the Astrophysical Journal
\renewcommand{\thefootnote}{\arabic{footnote}}
\begin{abstract}

Observations of Jupiter's gravity field by Juno have revealed surprisingly small values for the high order gravitational moments, considering the abundances of heavy elements measured by Galileo 20 years ago. The derivation of recent equations of state for hydrogen and helium, much denser in the Mbar region, worsen the conflict between these two observations. In order to circumvent this puzzle, current Jupiter model studies either ignore the constraint from Galileo or invoke an ad hoc modification of the equations of state. In this paper, we derive Jupiter models which satisfy both Juno and Galileo constraints. We confirm that Jupiter's structure must encompass at least four different regions: an outer convective envelope, a region of compositional, thus entropy change, an inner convective envelope and an extended diluted core enriched in heavy elements, and potentially a central compact core. We show that, in order to reproduce Juno and Galileo observations, one needs a significant entropy increase between the outer and inner envelopes and a smaller density than for an isentropic profile, associated with some external differential rotation. The best way to fulfill this latter condition is an inward decreasing abundance of heavy elements in this region. We examine in details the three physical mechanisms able to yield such a change of entropy and composition: a first order molecular-metallic hydrogen transition, immiscibility between hydrogen and helium or a region of layered convection. Given our present knowledge of hydrogen pressure ionization, combination of the two latter mechanisms seems to be the most favoured solution.

\end{abstract}


\section{Introduction}
For more than 30 years, guided by the observations of Voyager and Pioneer \citep{Campbell85}, all traditional models of Jupiter have been described as 2-layer or 3-layer models, namely a
homogeneous, convective gas rich envelope, generally split in a molecular/atomic outer part and an ionized/metallic inner one, and a supposedly solid core (e.g., \citet{Chab92,Saumon2004}), as first intuited by \citet{Stevenson1977_2, Stevenson1977}.
Later on, Galileo provided
new constraints on Jupiter's outer layers composition 
(\citet{vonzahn1998} for helium and  \citet{Wong2004} for the re-analysed results of the heavy elements). 
Finally, in 2017 and 2018, the observations of Juno reported in \citet{bolton2017}
and \citet{Iess2018} stressed the need to resolve a real puzzle: how to model an internal structure
of Jupiter matching both the observations of Galileo, revealing a highly supersolar outer element abundance, and Juno, for the gravitational moments ? 

The trouble indeed is to reconcile the low value of Juno's high order even gravitational moments, 
$J_4$ to $J_{10}$, and the high value 
of helium and heavy elements observed by Galileo, $\ygal$ and $\zgal$. The higher the order of a gravitational moment, 
the more sensitive it is to the outermost part of the planet. Hence, the most important 
physical parameters to determine the values of $J_4$ to $J_{10}$, for a given mass and $J_2$, 
are the abundances of helium and heavy elements in the external envelope of the planet.


In order to resolve this puzzle, \citet{Wahl17} had either to invoke an ad-hoc modification
of their H/He EOS or to reduce the outer heavy element content compared with Galileo's observations. 
\citet{Guillot2018} also
allowed the outer heavy element content
to vary from $0$ to $\zgal$, but their model matching all Juno $J_n$ values have an 
amount of heavy elements in the atmosphere which is not compatible with the Galileo constraints (Guillot, private com.).

In this paper, we present models of Jupiter
which do fulfill both Juno and Galileo observational constraints. 
We expose the method and 
the different physics inputs in \S\ref{sec:method}.
In \S\ref{sec:simple}, we
demonstrate the necessity to have several different regions in Jupiter's interior and show that traditional 2- or 3-layer models
fail to reproduce the observations. In \S \ref{sec:z_neg}, we show that a locally {\it inward decreasing abundance of heavy elements} in the Mbar region is the favoured solution to resolve this puzzle. We explore in details the possibility to have such an element distribution.

Our final models are presented in \S\ref{sec:final}. We first show that, without
the presence of a sharp entropy increase somewhere within the gaseous envelope, the values of
$J_6$ to $J_{10}$ are too large compared with Juno's ones, which implies to invoke an implausibly large amount of differential rotation. Indeed, a strong entropy increase in the region of hydrogen 
metallisation (around 1 Mbar) yields higher internal temperatures, 
allowing a larger amount of heavy elements in the central region (\S \ref{ssec:entropy}). 
This in turn affects the high order gravitational moments and enables us to derive Jupiter models which
satisfy both Juno and Galileo observational constraints.
We examine in detail the possible physical mechanisms leading to this type of internal structure and discuss their implications for the physics of hydrogen
pressure ionization. We also examine the possible amount of differential rotation in Jupiter. In \S\ref{sec:discussion}, we summarize and examine the validity of
the major physical assumptions that have been made throughout this study.
Section \ref{sec:conclusion} is devoted to the conclusion.

\section{Method}
\label{sec:method}
\subsection{Concentric MacLaurin Spheroids}
\label{ssec:CMS}

Our Jupiter models are calculated with the Concentric Maclaurin Spheroid method (\citet{Hub2012}, \citet{Hub2013}). As demonstrated in \citet{Debras2018}, in order to yield valid models, the method must
fulfill several mathematical and numerical constraints, in terms of {\it numbers and spacing} of the spheroids and of the treatment of the
outermost spheroids. Accordingly, the spheroids implemented in our 
calculations are spaced exponentially, their equatorial radius
is $\lambda_i = 1 - (\mathrm{e}^{i \beta} - 1)/(\mathrm{e}^{N \beta} - 1)$
with $N$ the number of spheroids, $i$ ranging from $0$ to $N-1$, $\beta = 6/N$ and 
the upper atmosphere is neglected\footnote{This implies an irreducible error of the order of $10^{-7}$  on $J_2$ and a few $10^{-8}$ on higher order moments, which is negliglible compared to the possible impact of 
differential rotation (\citet{Debras2018}, \citet{Kaspi2017})}. 
In this paper, we examine which kind of model
is compatible with Juno's observations, provided the difference can be explained by the {\it maximum} allowed amount 
of differential rotation, i.e. differential rotation penetrating down to 10,000 km (see \citet{Guillot2018} and \citet{Kaspi2017}). 
Said differently, we want the uncertainty on 
the $J$ values obtained for an acceptable model to be smaller than the one due to this maximum possible level of differential rotation. 
At this level, we checked that 512 spheroids yield a sufficient precision and that using 1000 spheroids or changing the $\beta$ parameter
does not significantly affect the conclusions. 
Deriving more precise models, fulfilling precisely all Juno's and Galileo's constraints with smaller levels of differental
rotation, however,
requires {\it at least} 1000 spheroids
to ensure that the discretisation error is negligible compared with the other sources of error on the evaluation
of the gravitational moments.
The various parameters used for Jupiter throughout this work are reported in Table \ref{values_Jupiter}.

\begin{table*}
\caption{Values of the planetary parameters of Jupiter.}
\label{values_Jupiter}
\centering
\begin{tabular}{ll}
\hline
\hline
Parameter & \multicolumn{1}{c}{Value} \\
\hline
$G$$^a$ (global parameter) & $6.672598 \times 10^{-11} \pm 2 \times 10^{-17}$ m$^3$kg$^{-1}$s$^{-2}$ \\
$G\times M_J$$^{b}$ &  ($126 686 533 \pm 2) $ $\times 10^9$m$^3$s$^{-2}$\\
$M_J$ & $1.89861 \times 10^{27}$ kg \\
$R_{eq}$$^c$ & $71492 \pm 4$ km \\
$R_{polar}$$^c$ & $66854 \pm 10$ km \\
$\omega$$^d$ & $1.7585324 \times 10^{-4}$ $\pm 6 \times 10^{-10}$ s$^{-1}$ \\
$\bar{\rho}$ & $1326.5 $ kg\,m$^{-3}$ \\
$m = 3 \omega^2/4 \pi G \bar{\rho}$ & 0.083408 \\
$q = \omega^2 R_{eq}^3/G M_J$ & 0.0891954 \\
$J_2 \times 10^6$$^e$ & $14696.572 \pm 0.014$ \\
$-J_4 \times 10^6$$^e$ & $586.609 \pm 0.004$\\
$J_6 \times 10^6$$^e$ & $34.198 \pm 0.009$\\
$-J_8 \times 10^6$$^e$ & $2.426 \pm 0.025$\\
$J_{10 } \times 10^6$$^e$ & $0.172 \pm 0.069$\\
\hline
\end{tabular}
\caption{$R_{eq}$ and $R_{polar}$
are observed at 1 bar. The value of the pulsation is chosen following
\citet{Archinal2011}.\\
(a) \citet{Cohen87}
(b) \citet{Folkner2017} (c) \citet{Archinal2011}
(d) \citet{Riddle76}
(e) \citet{Iess2018}
}
\end{table*}

\subsection{Equations of state}
\label{ssec:eos}

Throughout this work, we use for the H/He mixture a combination of
the new equation of state (EOS) recently derived by \citet{Chabrier2018}, based on semi-analytical
models in the low (molecular/atomic) and high (fully ionized) temperature-density domains and quantum molecular dynamic (QMD) calculations
in the intermediate pressure dissociation/ionization regime, and the \citet{MH13} equation
of state that takes into account non-ideal correlation effects.
As in \citet{Miguel2016}, we have first calculated
a pure H table by calculating, at each $P$-$T$ point:

\begin{eqnarray}
\frac{1}{\rho_{\mathrm{MH13}}} &=& \frac{X_\mathrm{MH13}}{\rho_H}+ \frac{Y_\mathrm{MH13}} {\rho_{\mathrm{He},\mathrm{New}}} 
 \Rightarrow  \frac{1}{\rho_{\mathrm{H}}}=\frac{1}{X_\mathrm{MH13}}\left(\frac{1}{\rho_{\mathrm{MH13}}} - 
\frac{Y_\mathrm{MH13}} {\rho_{\mathrm{He},\mathrm{New}}} \right), \label{new1} \\
S_\mathrm{MH13} &=& X_\mathrm{MH13} S_\mathrm{H} + Y_\mathrm{MH13}S_{\mathrm{He},\mathrm{New}} +  S_\mathrm{mix}
\Rightarrow   S_H =  \frac{1}{X_\mathrm{MH13}}\left(S_\mathrm{MH13} - Y_\mathrm{MH13}S_{\mathrm{He},\mathrm{New}} - S_\mathrm{mix}\right),
\label{new2}
\end{eqnarray}
where $\rho_H$ is the mass density for pure hydrogen, $\rho_{\mathrm{MH13}}$ the density derived from MH13 by spline procedures, 
$\rho_{\mathrm{He},\mathrm{New}}$ the helium density in the \citet{Chabrier2018} EOS, 
$S_H$ the sought pure hydrogen specific entropy, $S_\mathrm{MH13}$ the splined specific entropy from MH13,
$S_{\mathrm{He},\mathrm{New}}$ the helium specific entropy in the new EOS, all at the same ($P$,$T$), and $X_\mathrm{MH13}=0.7534$, $Y_\mathrm{MH13}=0.2466$ 
the mass fractions of hydrogen and helium in the MH13 simulations. Finally, $S_\mathrm{mix}$ is 
the mixing specific entropy defined as: 

\begin{eqnarray}
\frac{S_\mathrm{mix}}{k_b} = \frac{1}{M_{\mathrm{H},\mathrm{He}}} \Bigl[ N_\mathrm{H} \ln \left(1 + \frac{N_\mathrm{He}}{N_\mathrm{H}} \right)
+ N_\mathrm{He} \ln \left(1 + \frac{N_\mathrm{H}}{N_\mathrm{He}} \right) \Bigr], 
\label{new3}
\end{eqnarray}
with $N_\mathrm{H}$ and $N_\mathrm{He}$ the numbers of $H$ and $He$ particles, respectively, of number fractions $x_i=N_i/(N_H+H_{He})$ ($i\equiv$H or He),
$M_{\mathrm{H},\mathrm{He}}={\bar A}m_H$ the total mass of H+He, ${\bar A}=\sum_i x_iA_i$ the mean atomic number and $m_H=1.660\times 10^{-27}$ kg the atomic mass unit. 
This "mixed" (Chabrier et al./MH13) pure hydrogen EOS is then combined with the new pure helium one (\citet{Chabrier2018}, Soubiran et al. (in prep)) to obtain a complete EOS for the H/He mixture at any given helium mass fraction $Y$.

\begin{figure}[ht!]
\center
\includegraphics[width=0.5\textwidth]{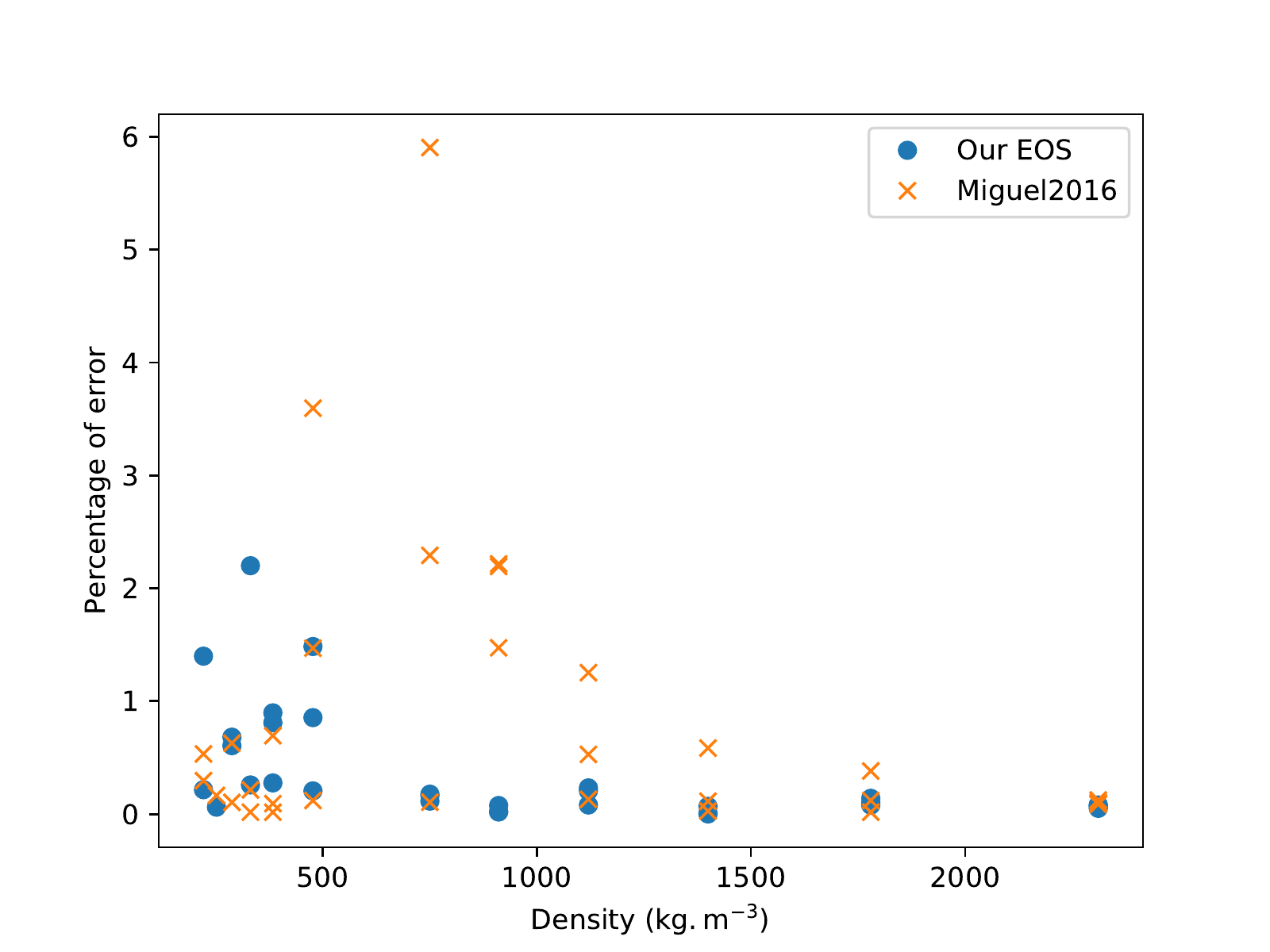}
\caption{Relative error on the density between MH13 and our or \citet{Miguel2016} mixed EOS for Y= 0.2466 .\label{fig:diff_eos} }
\end{figure}

Figure \ref{fig:diff_eos} displays the relative error on the density between our or the \citet{Miguel2016} EOS
and the MH13 one, $(\rho-\rho_{MH13})/\rho_{MH13}$, for Y= 0.2466, the helium fraction used in MH13. For \citet{Miguel2016}, we 
have combined their published pure H table with a He table from SCvH with a cubic order spline. 
The comparisons are made for 32 $(T,P)$ points from MH13 corresponding to an entropy characteristic of Jupiter interior, 7-8 $k_\mathrm{B}/$proton. These points are used as inputs in our or \citet{Miguel2016} mixed EOS to calculate the corresponding density and entropy, to be compared with the MH13 one. 
As seen in the figure, above $500\, \mathrm{kg\,m}^{-3}$, the pressure ionization domain in Jupiter, the difference between our and
MH13 results is always $<0.5 \%$,
which is less than the numerical error in MH13, whereas for the \citet{Miguel2016} EOS the differences are significant, a major issue
in the present context where a very accurate density profile is required to derive reliable gravitational moments.

For the heavy elements in the H/He rich envelope, composed essentially of volatiles (H$_2$O, CH$_4$, NH$_3$), we use a recent EOS for water, based on QMD simulations at high density (\citet{LicariPhD}, \citet{Mazevet2018}). Here also, this EOS has been shown to adequately reproduce available
Hugoniot experiments \citep{Knudson2012}.
Given the small number fraction of CH$_4$ and NH$_3$ compared with H, He and even H$_2$O, we do
not expect the assumption to generically treat all the volatiles with the water EOS to be consequential on the results. When considering a {\it diluted} core, we combine the above water 
EOS with the the Sesame "drysand" one \citep{drysand} to take into account additional heavy elements such as silicates and iron. Finally, when including a central {\it compact} core, we use
a 100\% "drysand" EOS. We verified that using e.g. an EOS for pure iron instead
of the drysand one for this region does not make noticeable difference. Note that ab initio simulations have shown that, under the characteristic temperature and pressure conditions typical of Jupiter deep interior, water is in a liquid state and is fully soluble in metallic hydrogen \citep{Wilson2012}. Similarly, solid SiO$_2$ and MgO, representative examples of planetary silica and rocky material, are also found to be soluble in H$^+$ under similar conditions (\citet{Gonzalez2014}, \citet{Wilson2012_PRL}, \citet{Wahl2013}). These thermodynamic considerations support a core erosion for Jupiter typical central conditions and thus a mixed $H/He/Z$ eos in such a region.

As shown by \citet{Soubiran2016}, the inclusion of 
heavy elements in a H/He/Z mixture under Jupiter-like internal temperature and density conditions can be performed with the so-called Additive-Volume-Law (AVL) provided we use an 
\textit{effective} volume (density) for the heavy species. We verified that our EOS 
is consistent with the work of \citet{Soubiran2016}, and hence that oubetween at most 0.1 and 0.3 Mbar water EOS, as representative of volatiles in Jupiter, can be used throughout the entire $T$-$P$ domain from Jupiter's atmosphere to the center.


For  H/He/Z mixtures, our EOS are then combined at each given 
$(P,T)$ point
throughout the AVL :

\begin{eqnarray}
\frac{1}{\rho} = \frac{1-Z_\mathrm{water}-Z_\mathrm{drysand}}{\rho_{\mathrm{H},\mathrm{He}}} +
\frac{Z_\mathrm{water}} {\rho_\mathrm{water}} + 
\frac{Z_\mathrm{drysand}} {\rho_{\mathrm{drysand}}} \hskip1.cm {\rm at} \,\,(P,T)={\rm constant},
\end{eqnarray}
where $\rho_{\mathrm{H},\mathrm{He}}$, $\rho_\mathrm{water}$ and $\rho_{\mathrm{drysand}}$ are 
the densities of the H/He mixture, water and drysand, respectively, 
and $Z_\mathrm{water} = M_\mathrm{water}/M$, $Z_\mathrm{drysand} =M_\mathrm{drysand}/M$
the mass fractions of water and drysand, respectively, with $M$ the mass of the planet.
Note that the accuracy of the AVL for the hydrogen/water mixture under the relevant conditions for Jupiter interior has been verified 
with QMD simulations \citep{Soubiran2015}.

Given the small number fraction of heavy elements compared with H and He, the $P$ and $T$ used to 
calculate the densities in the H/He/Z mixture are the ones obtained with
the H/He mixture only. Similarly, the entropy of heavy elements 
can be neglected (see \cite{Soubiran2016}) and even when their mass fraction becomes $Z \gtrsim 0.2$, they
affect the total mixing entropy by at most 2$\%$, which
represents a few per thousands of the total entropy. Moreover, 
this occurs only in the deepest
part of the planet, with little impact on the gravitational moments.
Hence, the total entropy
is evaluated as the entropy of a pure H-He mixture 
with effective hydrogen and helium mass fractions $X^\mathrm{eff}=X/(1-Z)$ and $Y^\mathrm{eff}=Y/(1-Z)$, respectively, with $X^\mathrm{eff} +Y^\mathrm{eff}=1$,
and $X=M_H/M$,
$Y=M_{He}/M$, and $Z = M_Z/M=Z_\mathrm{water}+Z_\mathrm{drysand}$.
\\

\subsection{Galileo constraints on the composition}
\label{Gal}

For the outer element abundances, the observations of Galileo give 
\begin{gather*}
\dfrac{\ygal}{ (\xgal + \ygal)} =0.238\pm 0.005, \ \, \\
\zgal = 0.0167\pm 0.006,
\end{gather*}
where $\xgal$ and $\ygal$ are the
observed mass abundances of hydrogen and 
helium, respectively. $\zgal$ is the abundance of heavy elements in the high envelope 
measured by Galileo, but the real abundance of heavy elements should be larger. This 
implies that $\ygal$ and $\xgal$ are only defined relatively to each 
other and that $\xgal + \ygal + \zgal$ can be larger than 1. 
In all the following models of this paper, except if stated otherwise,
we impose the external atmosphere to have helium and heavy element mass fractions: 

\begin{gather*}
Y_\mathrm{ext} = 0.23 \\
Z_\mathrm{ext} = 0.02,
\end{gather*}
which corresponds to $Y_\mathrm{ext}/(X_\mathrm{ext}+Y_\mathrm{ext}) = 0.2347$. 
As just mentioned, this $Z$ value is most likely a 
lower limit for the heavy element content in the external envelope of Jupiter (see \S\ref{ssec:high_Z} for a detailed exploration of this issue). 
Forgetting Galileo's constraints, i.e.
reducing the observed amount of heavy elements, drastically reduces the constraints on the models and allows the derivation
of a large range of models compatible with Juno's data. Relaxing these constraints thus drastically simplifies the calculations of models consistent with {\it only} Juno's observations. Such simplifications, as done in all recent studies (\citet{Guillot2018}, \citet{Wahl17}, except when using an ad-hoc modification
of their EOS), however, can hardly be justified (see \S \ref{sec:discussion}). In all our calculations, the planet's mean helium fraction 
is fixed to the protosolar value: $\ymean/(\bar X + \bar Y) = Y_\odot=0.275$ (see e.g., \citet{Anders1989}).

\section{Simple benchmark models}
\label{sec:simple}

In this section, we show that traditional homogeneous, adiabatic 2 or 3 layer interior models for Jupiter are excluded by the new observations
and that the planet must consist of several different regions. 

\subsection{Homogeneous adiabatic gaseous envelope}

\begin{figure}[ht!]
\centering
  \includegraphics[width=.5\linewidth]{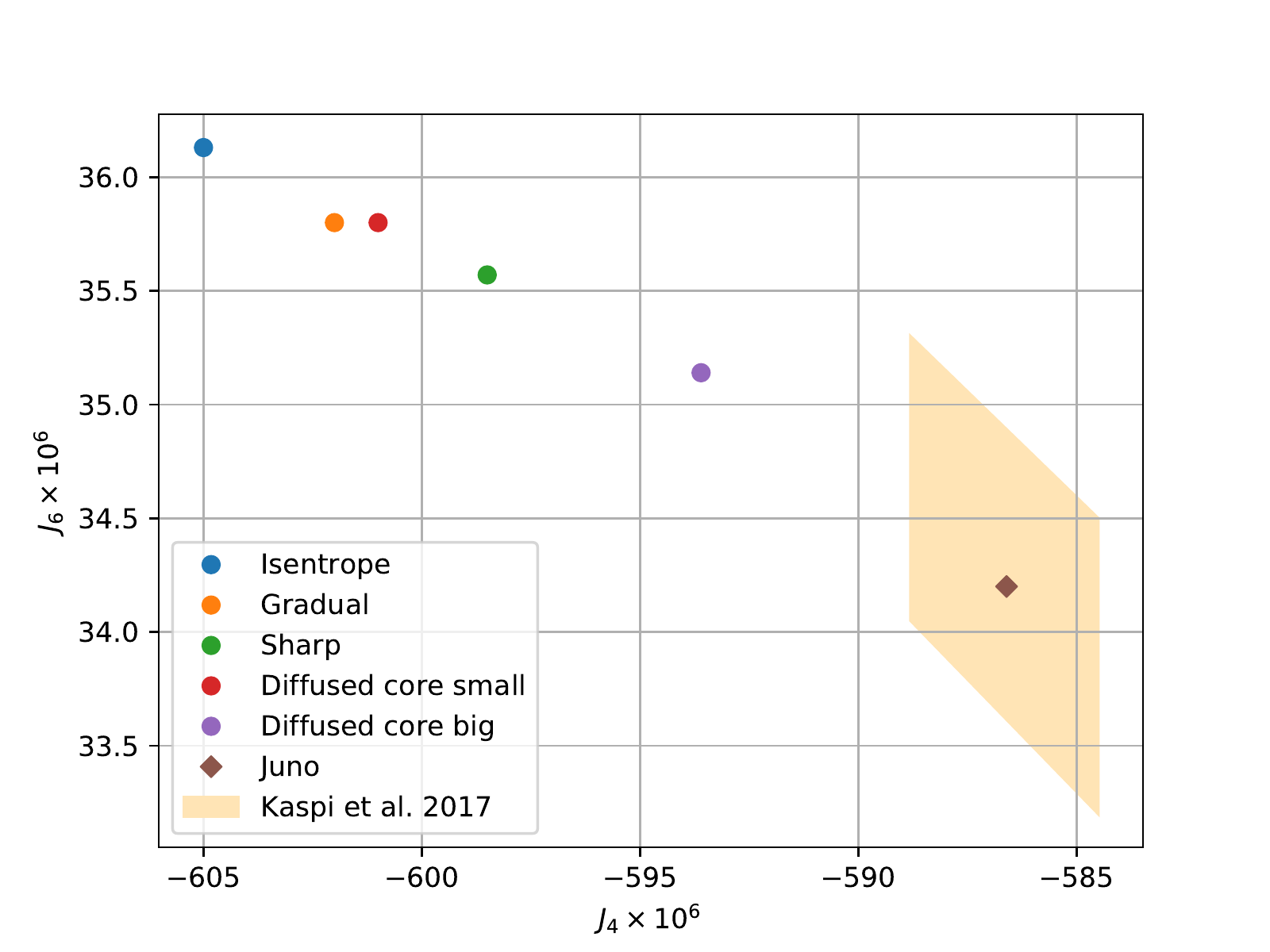}
\caption{$J_4 \times 10^6$ vs $J_6 \times 10^6$ for different models (see text) and 
Juno's values \citep{Iess2018} with the error bars too small to be 
seen on this Figure. The shaded area correspond to the uncertainty 
on the graviational moments arising from differential rotation shallower
than 10000 km, as evaluated by \citet{Kaspi2017}.}
\label{fig:isentrope}
\end{figure}

We first calculate an
isentropic model, composed of one homogeneous convective isentropic gaseous envelope, with $\ymean=0.275$, and a spherical compact core of constant density. 
The total heavy material content 
is determined 
to obtain the correct mass of the planet. This model
reproduces the $J_2$ observed value within $10^{-7}$, 
the maximum intrinsic precision of the CMS method \citep{Debras2018}. 
The $J_4$ and $J_6$ values are compared to Juno's ones in 
Figure \ref{fig:isentrope} under the labels "Isentrope".
The differences between the observed and calculated values are about $3 \%$ and $6 \%$, respectively, 
well above any numerical source of error.

The only possibility to reconcile the observed and theoretical values would be the presence of strong differential 
rotation affecting the calculation of the gravitational moments, not implemented 
in the CMS calculations. However, the results of \citet{Kaspi2017} show
that with various flow profiles extending more than $10 000$ km within the planet, 
the change in $J_4$ is at most $0.7 \%$ (see their Figure 4). 
Furthermore, the study of \citet{Cao2017} excludes such a deep differential rotation.

\begin{itemize}
\item This yields the first robust conclusion: {\it Jupiter's interior is not isentropic}. 
\end{itemize}

\subsection{A region of compositional and entropy variation within the planet}
\label{ssec:composition_change}

The next step in increasing complexity is to change the composition, then the entropy, somewhere in the planet.
There are two physically plausible domains: a diluted core extending throughout a substantial fraction of the interior and/or a 
region of either layered convection, hydrogen metallisation or H/He phase separation somewhere within the envelope.
 
\subsubsection{Two possible locations}
\label{sssec:location}

A diluted core or a region of layered convection further up in the planet can emerge
during the evolution of the planet or can be inherited from the formation process (see e.g. \citet{Stevenson1985}, \citet{Chabrier2007} and
references therein, \citet{Helled2017}) and is characterised by a compositional gradient $\nabla \mu=(\partial \mu/\partial r)\propto \nabla Z=(\partial Z/\partial r)$ with $\mu$ the mean molecular weight, thus an entropy
gradient $\nabla S=(\partial S/\partial r)$. For hydrogen metallisation, 1$^{st}$-principle numerical simulations suggest it could occur through a first-order phase transition (usually denominated
plasma phase transition, PPT, or liquid-liquid transition, LLT) in a domain $P_c\sim 1$-2 Mbar, for temperatures below the critical temperature $T\le T_c\simeq 2000$-5000 K (\citet{Morales2010} , \citet{Lorenzen2011}, \citet{Morales2013}, \citet{Knudson2015}, \citet{Mazzola2018}).
Experiments on liquid deuterium, D$_2$, seem to be consistent with these figures, even though significant differences still persist between
various experiments (see e.g. \citet{Knudson2015}, \citet{Celliers2018}). Until this issue is resolved definitively, an entropy discontinuity, $\Delta S$ due to hydrogen pressure ionisation in Jupiter's envelope, although unlikely, can thus not be definitely excluded. In a similar vein, H/He phase separation, also a first-order transition, will also yield an entropy discontinuity, provided the local temperature is lower than the critical one for the appropriate He concentration (see below). Last but not least, a
regime of double-diffusive layered convection could develop somewhere within the planet interior, triggered either by one of these two transitions (or
by any phase separation involving some heavy component insoluble in metallic hydrogen) and/or simply by a local compositional gradient \citep{Leconte2012}. Phase transitions, indeed, notably endothermic ones, are suspected of enforcing
layered convection, for instance in the Earth's mantle (Schubert et al. 1975, \citet{Christensen1985}). The physical reason
is the release of latent heat at the transition, which leads to thermal expansion and temperature advection which tend to hamper convection.
It is interesting to note that, due essentially to the larger entropy in the plasma phase than in the molecular one, a PPT is an endothermic
transition, i.e. $dP/dT<0$ along the transition critical line, according to the Clausius-Clapeyron equation, $dP/dT\propto -\Delta S/\Delta \rho$. As for the H/He immiscibility, ab initio simulations, while still differing substantially, seem to suggest
a critical temperature for $x_{He}=0.08$ in the range $\sim 2000$-800{}0 K for $P\gtrsim 1$ Mbar (100 GPa), with a weak dependence upon pressure in the $T,P$ domain relevant for Jupiter, suggesting $dP/dT\sim 0$ for the protosolar helium value (\citet{Lorenzen2009}, \citet{Morales2009},
\citet{Morales2013_2}) (see Fig. 13 below). 
 
Therefore, the entropy variation in the gaseous envelope could occur either within a region of layered convection due to compositional gradients or because of either a PPT or a H/He phase separation.
Needless to say, not only these three physical processes are not exclusive but they are likely
to be tightly linked and thus to take place in the same more or less extended hereafter denominated "{\it metallization boundary region}" near the Mbar.

\subsubsection{Results}

\label{entropy}

\begin{figure*}[ht!]
\begin{subfigure}{.5\linewidth}
  \flushleft
  \includegraphics[width=\linewidth]{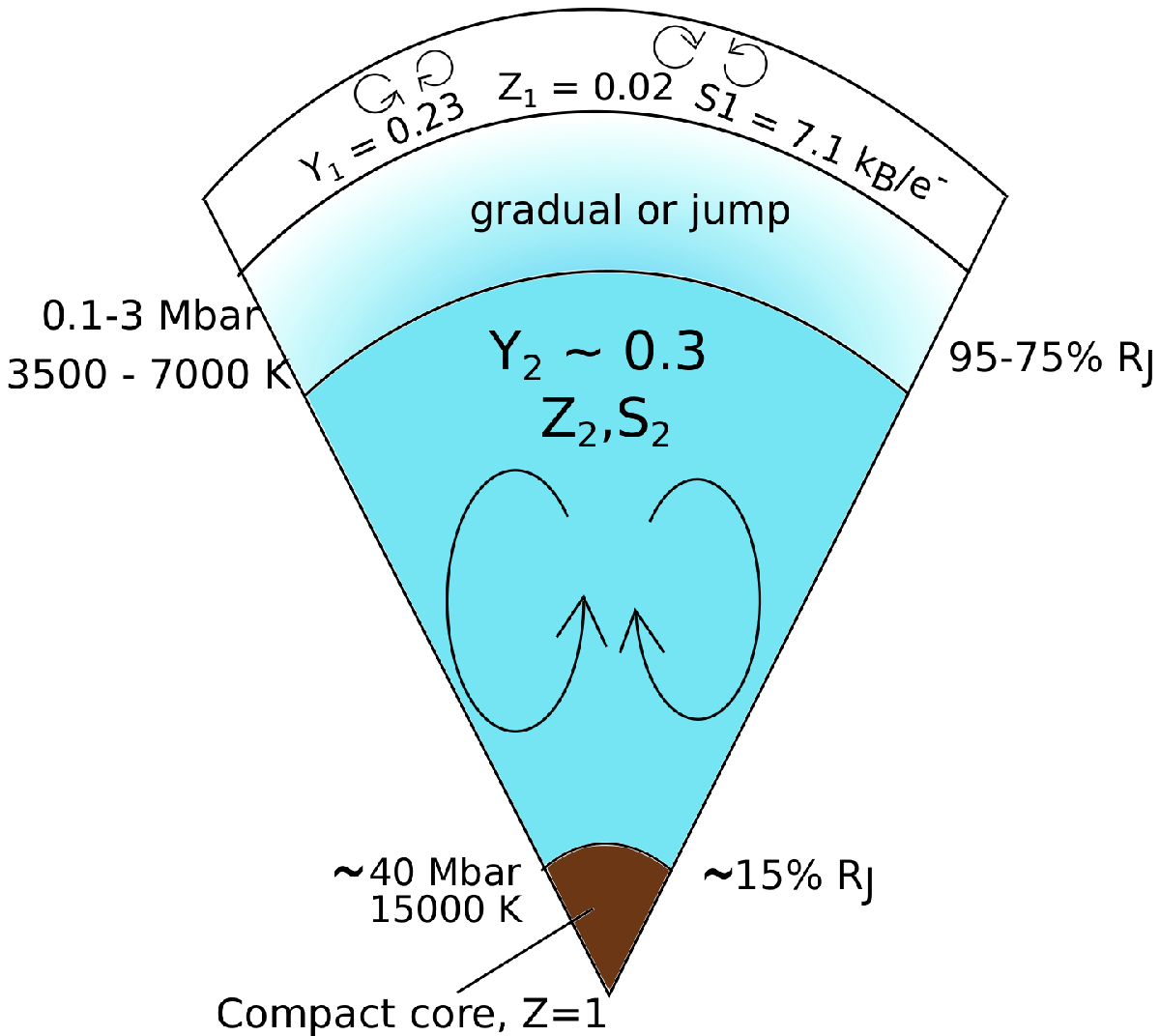}
  \caption{}
  \label{fig:simple_1}
\end{subfigure} 
\begin{subfigure}{.5\linewidth}
  \flushright
  \includegraphics[width=\linewidth]{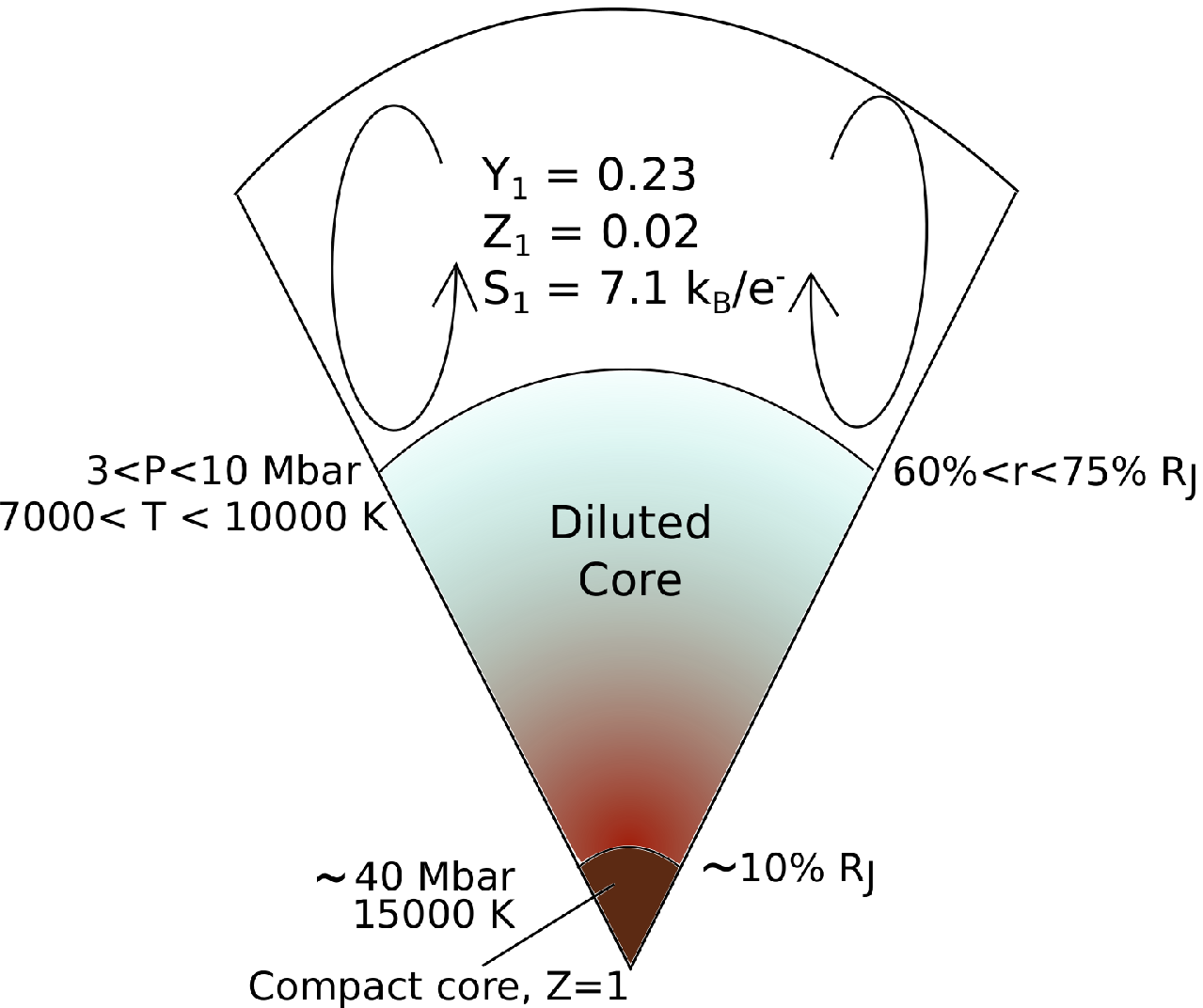}
  \caption{}
  \label{fig:simple_2}
\end{subfigure}
\caption{Simple structures of Jupiter with an internal entropy variation either in the envelope (left) or in the core (right) (see \S\ref{entropy}). None of these models can match the observations
of both Juno and Galileo.\label{fig:simple}}
\end{figure*}

Following up on the above analysis, we have explored two types of models with entropy and compositional changes {\it either in the central region (the "diluted core") or in the gaseous envelope}, as schematically illustrated in Fig. \ref{fig:simple}. In case of a diluted core unstable to double diffusive behaviour, \citet{Moll2017} showed that 
a central seed could survive to erosion longer than the lifetime of Jupiter. The fact that, under Jupiter 
central $T$ and $P$ conditions, iron and silicates are under a solid form (see e.g., \citet{Musella2018}) tends to favor the presence
of such a central seed. We thus consider the presence of a compact core at the very center of the planet.
For the entropy variation within the gaseous envelope, we have considered either an abrupt ($\Delta S$) or a gradual ($\delta S$)
change in entropy and composition at the metallisation boundary, between at most 0.1 and 3.0 Mbar. 
The obtained $J_4$ and $J_6$ values are given in Fig. \ref{fig:isentrope}. 
For the models with a change of composition in the gaseous envelope, the values for a gradual 
or a sharp entropy change are plotted under the labels 'Gradual'
and 'Sharp', respectively.

In case of a gradual (continuous) change, which implies a continuous 
molecular H$_2$ to metallic H$^+$ transition and no H/He phase separation, the smooth change in entropy is simply due to a composition change (see eqns. (2)-(4)).
In case of a first order
molecular-metallic transition, the abrupt change in entropy $\Delta S$ is used as a free parameter, discretised over a certain number of spheroids to get the proper $J$ values.

In all cases, none of these two types of models,  whatever the type of change of composition and entropy, sharp or gradual, was found to be able to yield $J_4$  and $J_6$ values sufficiently close to the observed ones (labeled 'Juno') to be explained
by differential rotation or deep winds.
\\

Another "simple" possible interior structure model is the one suggested by
\citet{Leconte2012}: the entire planet would be made of alternating
convective and diffusive layers. 
These authors, however, pointed out that this
entire double-diffusive interior model could be replaced by a model with a localised double-diffusive buffer in the envelope, surrounded by large scale convective envelopes (see their \S4.3), similar to the type of model explored above, which is found to be excluded. We will return to this point in \S \ref{ssec:entropy} and \ref{sec:discussion}.


This yields the following conclusions:
\begin{itemize}
\item {\it Conclusion 1}: The first conclusion of this section is that models of Jupiter 
displayed in Figure \ref{fig:simple}(a) and \ref{fig:simple}(b)
cannot fulfill both Juno and Galileo observational constraints. 
One needs a mix of these two types of models: Jupiter is at least composed of an envelope 
split in two parts (an outer molecular/atomic envelope with Galileo element composition 
and an inner ionized one) separated by a region of compositional change, 
{\it and} a diluted core extending throughout a significant fraction of the planet. A compact 
core can also be present at the center of the planet.

\item{\it Conclusion 2}:
In order to decrease the values of $|J_4|$ and $|J_6|$ in the models of Figure \ref{fig:simple}, 
we realised that either $\Delta S$ had to be substantial (an issue explored in \S \ref{sec:final}) or {\it the heavy element content must decrease with depth in the outer
part of the planet}. This local decrease of $Z$ is balanced by an increase of $Y$ so that the density
(and the molecular weight, see \S\ref{sec:discussion}), of course, increases with depth.
As mentioned in the introduction, one of the most stringent constraints on the models are the $Y$ and $Z$ values observed by Galileo, 
which are surprisingly high for the observed values of the high order gravitational moments. {\it A local inward decrease of the metal content
in some region of the planet's gaseous envelope appears to be the favoured solution to resolve this discrepancy.}
This is examined in detail the next Section.
\end{itemize}

It is essential at this stage to stress the crucial role played by the H/He EOS. 
Using the SCvH EOS, \citet{Chab92} and subsequent similar models, which fulfilled Galileo and Voyager observational constraints,
could relatively easily fulfill as well the Juno ones. This is entirely due to the 
SCvH EOS (or, similarly, the R-EOS one \citep{Nettelmann2012}, see Fig. 11 of \citet{MH13} or Fig. 27 of \citet{Chabrier2018}): for a given entropy, such an EOS has a lower density (pressure) in the $\sim$Mbar region than our new EOS,
enabling a larger amount of heavy element repartition in Jupiter interior, relaxing appreciably the constraints on possible models. The constraints become much more stringent with stiffer, more accurate EOS's.

\section{Locally inward decreasing Z-abundance in the gaseous envelope}
\label{sec:z_neg}

\subsection{Inward decreasing abundance of heavy elements in some part of the outer envelope}

Two physical processes can lead to a locally decreasing abundance of heavy elements with depth in Jupiter's outer envelope, 
i.e. a locally positive gradient $\nabla Z>0$: a "static" one,
based on thermodynamic stability criteria, and a "dynamic" one, which involves non-equilibrium processes.
For sake of completeness, a third, more "exotic" process, involving external events, is also discussed.

\subsubsection{Thermodynamic stability}
\label{thermo}

\citet{Salpeter1973} and \citet{Stevenson1977} (see also \citet{Stevenson1979}) suggested the occurence of helium differentiation in giant planet interiors, 
either in the same or in a different region that the H$_2$-H$^+$ pressure metallisation of 
hydrogen. These authors suggested that minor constituents, namely the heavy elements, 
could suffer differentiation in a similar or even larger way as helium. Unfortunately, 
phase diagram calculations of two or more components under the typical relevant conditions for Jupiter (about 1 Mbar and 5000 K)
are scarce, or even inexistent, so finding out which element, under which molar concentration, 
prefers the H$^+$-rich or He-rich phase remains to be determined. The only existing study is the one by \citet{Wilson2010}.
Ab initio simulations by these authors suggest that Ne association with He is thermodynamically favoured, while the opposite is true for Ar,
which is found to be more soluble with H$^+$. The underlying physical reason is the argon atom additional electron shell which increases
its effective volume with respect to He due to the Pauli exclusion principle. If this explanation is correct, Kr and Xe should likewise be
soluble in metallic hydrogen, which is consistent with their observed nondepletion in Jupiter's atmosphere.

It is indeed intuitively appealing to think that in case some species, $Z_i$, is pressure ionized, it might become immiscible with neutral helium, as for H$^+$/He,
 due to the strongly repulsive pseudo-potential, as in the case e.g. of alkali metals \citep{Stevenson1979}.
For some element (atom or molecule) to differentiate in the midst of some mixture, one needs its interaction energy in
the mixture, typically the molecule or electron binding energy, to be larger (in absolute value) than the ideal mixing entropy, $-k_B\,T\ln\, x_Z$. Since
the most abundant heavy elements have a number fraction $x_Z \simeq 0.1\%$, this yields near the metallization boundary, $\sim 5000 $ K, $|E_b|\gtrsim 0.5$ eV, a condition rather easy to fulfill.
As mentioned above, all heavy elements, however, do not necessarily behave similarly.
Heavy noble gases, indeed, are more likely to form compounds (\citet{Hyman1964} \citet{Blackburn1966} \citet{Wilson2010}), suggesting that species like
neon, acting like helium, 
and argon, have a different behaviour in the H$^+$/He mixture.

In case of element differentiation, according to the Gibbs phase rule, $x_{II}=x_I\exp \{-\Delta G(P,T)/k_bT\}$, where $x_i$ denotes the number abundance of a given
species in phase  $I$ or $II$ (H$^+$-rich/poor, conversely He-poor/rich in the present context) and $\Delta G$ is the excess mixing enthalpy in the
mixture, the differentiation of a given heavy element can be similar or opposite to that of helium, 
yielding an increasing (resp. decreasing) abundance with depth in the former (resp. latter) case. In all cases, this yields a gradient of abundance $\nabla Z$ within some part of the planet envelope, with $\nabla Z>0$ for some heavy elements.


If H-He immiscibility, leading to a 
depletion of helium in the outer envelope, is triggered by the 
metallisation of hydrogen, the fact that hydrogen
metallisation in a H/He mixture is found to occur at lower pressures with decreasing helium fraction (e.g. \citet{Mazzola2018}) implies
that the pressure range of immiscibility
will extend with time because of both the planet's decreasing internal temperature and the decreasing abundance (depletion) of helium in the upper layers. 
The region of immiscibility can thus be relatively broad in Jupiter's interior, depending on when it started.
Note in passing that, if H/He differentiation occurs and is at least partly responsible for the redistribution of heavy elements in Jupiter's envelope, this excludes the H/He diagrams suggested by \citet{Morales2009} and \citet{Schottler2018} which predict no immiscibility within present Jupiter. This point will be discussed in detail in \S\ref{HHe}.

It should be stressed that, given the small number abundances of helium and heavy elements in Jupiter interior, whereas the aforedescribed demixing processes could lead to some
Z-enrichment in the planet outer envelope, consistent with or even larger than Galileo's observed value, this enrichment will remain modest and could not explain values significantly larger than $\zgal$. In this latter case, external acccretion seems necessary, as examined in \S\ref{ssec:accretion} below.

\subsubsection{Upward atomic motions}
\label{sssec:1st_order_met}

As explored thoroughly by \citet{Stevenson1977} in the case of a first-order hydrogen pressure ionization (PPT) and a
H/He phase separation occuring in the same region, the following process might occur.
At the onset of hydrogen metallisation, characterised by a pressure $P^+$, latent heat release will lead to the superposition of a overheated  (resp. supercooled) H$^+$-rich (resp. H$_2$-rich) layer, thus less (resp. more) dense than the surrounding medium, underneath (resp. above) the metallization boundary (see Fig. 2 of \citet{Stevenson1977}). Under such conditions, nucleation of bubbles might occur.
Concomitantly, He atoms will differentiate from H$^+$. If such H$^+$-rich/He-poor bubbles form, 
they will absorb heat by thermal diffusivity and will be lighter than the surrounding gas. 
The bubbles will then rise by buoyancy, up
to a pressure $P$ less than the metallisation pressure, $P<P^+$. They will then break and H$^+$ will recombine to form H$_2$, 
depleting little by little the upper envelope in He 
by mixing this convective region with H$^+$-rich bubbles while  enriching the lower envelope in He. 
Consequently, the heavy elements which, for chemical and/or thermodynamic reasons, have a preference 
for these H$^+$-rich/He-poor bubbles, rather than for the He-rich/H$^+$-poor surrounding medium, will 
be transported upwards and be depleted little by little in the deep envelope whereas the opposite will be true for species favoring 
association with helium atoms. 
Somehow, this is similar to an ongoing distillation process in the sense that the redistribution of elements arises from a
physical separation rather than a chemical reaction and mass is not {\it locally} conserved beneath the uppermost convective envelope. This occurs only if the heavy elements do not affect significantly the density of 
the bubbles, which must remain lighter than the surrounding gas. 
In the typical conditions of Jupiter's outer envelope, there are about 
$500$ times more atoms of hydrogen than of heavy elements. Therefore, for a typical atomic weight ratio ${\bar A}_Z/A_H\sim 15/1$ (average
between C, N and O atoms), such a process is possible. Heavier molecules (such as iron) being even more rare 
compared to H, gravitational considerations are still consistent with this scenario. 

Such a scenario has further theoretical support. First, noble gases have been known to be 
almost insoluble in metals since the end of the 19th century (\citet{Ramsay1897}
or \citet{Blackburn1966} for a review, and \citet{Wilson2010} for the case of neon and
argon). On the other hand, at high 
pressure, hydrogen can form complex polyhydrides molecules very efficiently, with 
many different atoms (sulfur, lithium, sodium, iron, ...; see e.g. \citet{Ashcroft2004} or \citet{Pepin2017}).
Therefore, at metallisation, non inert heavy elements tend to form 
polyhydrides within metallic H$^+$-rich bubbles. If, as discussed above, the density of these bubbles is less than the one of the surrounding
gas, these heavy elements will be transported upwards, enriching Jupiter's outer envelope while depleting the inner 
one. 
The formation of polyhydrides, however, has been probed experimentally so far up to $\sim 1500$ K \citep{Pepin2017} and remains to be explored  up to $T \sim 4500 K$,
the onset of H metallisation in Jupiter.
Further numerical or experimental work on the formation of polyhydrides at high pressures and temperatures would help assessing the validity of this process. 

Concomitantly with hydrogen metallization, and the formation of polyhydrides, we also expect reduction-oxydation (redox) reactions to occur. 
The loss of its $1s$ electron at hydrogen metallization makes H$^+$ prone to react with other heavy elements through electron transfer.  
The H$^+$ bubbles could then trap e.g. N, O or other elements, participating also
to en enrichment (resp. depletion) of these elements in the upper (resp. lower) envelope.
We recall, however, that in the absence of dynamical variations such as gravity waves or upward plume penetrations, the amount of overheating due to a PPT would be insufficient to yield homogeneous nucleation \citep{Stevenson1977}.

In all cases, the afore "rising bubble" process requires hydrogen molecular-metallic transition to occur through a first-order transition, leading to a local release of latent heat. As mentioned in \S \ref{sssec:location}, although a PPT is indeed found in some modern ab initio numerical simulations, its critical temperature remains to be determined
precisely, being predicted within the range $2000 \lesssim T_c \lesssim 5000$ K for a critical pressure $1 \lesssim P_c \lesssim 2$ Mbar.
Interestingly enough, 
a critical temperature $T_c\simeq 5000$ K around $\sim 1$ Mbar would be consistent with a PPT in the outer part of Jupiter. 
In the absence of a PPT in the envelope, H$^+$/He phase separation can possibly still occur but the outer observed oversolar abundance
of heavy elements can not be due to upward bubble motions. Phase separation of these elements with He, as discussed in \S\ref{thermo}, will thus be the favoured explanation.

\subsubsection{Accretion}
\label{ssec:accretion}

Finally, the overabundance of heavy elements in Jupiter's upper envelope can have a third explanation, namely one
or several giant impacts \citep{Iaro2007} or, similarly, ongoing accretion of planetesimals (e.g., \citet{Bezard2002}). This scenario, however, implies
that global internal convective motions must be inhibited somewhere in Jupiter,
 preventing the extra accreted material
to be redistributed homogeneously throughout the planet. Indeed, a global $Z$ abundance throughout the planet equal to the Galileo value would yield low-order
gravitational moments  
inconsistent with observations.
If convection inhibition is due to H/He immiscibility,
this latter must already have started when the external event took place. This in turn puts an important
constraint on the H/He phase diagram, notably on the critical $P, T$ values for $x_{He}=0.08$, Jupiter
helium protosolar concentration.  If inhibition is due to hydrogen metallization,
it implies that this latter must be a 1$^{st}$-order phase transition (yielding an entropy jump). 
Convection can also be inhibited by the onset of double diffusive convection, either as an enhanced diffusive process (oscillatory convection) or as layered convection, a process possibly triggered by extensive planetesimal accretion
(e.g. \citet{Stevenson1985}, \citet{Chabrier2007} and references therein) and/or by deposition of high entropy material onto the growing planet (Berardo \& Cumming 2017), preventing homogeneization of the envelope composition. 
In order to explain a genuine abundance significantly larger than Galileo's observed value, $Z_{ext}\simeq 2.5\times Z_\odot$,
the total accreted mass must be $M_\mathrm{excess}\lesssim 1.5\,\mearth$, a significant but not unplausible value.

\subsection{Constraints from the evolution}
\label{evol}

\begin{figure}[ht!]
\centering
\includegraphics[width=0.5\textwidth]{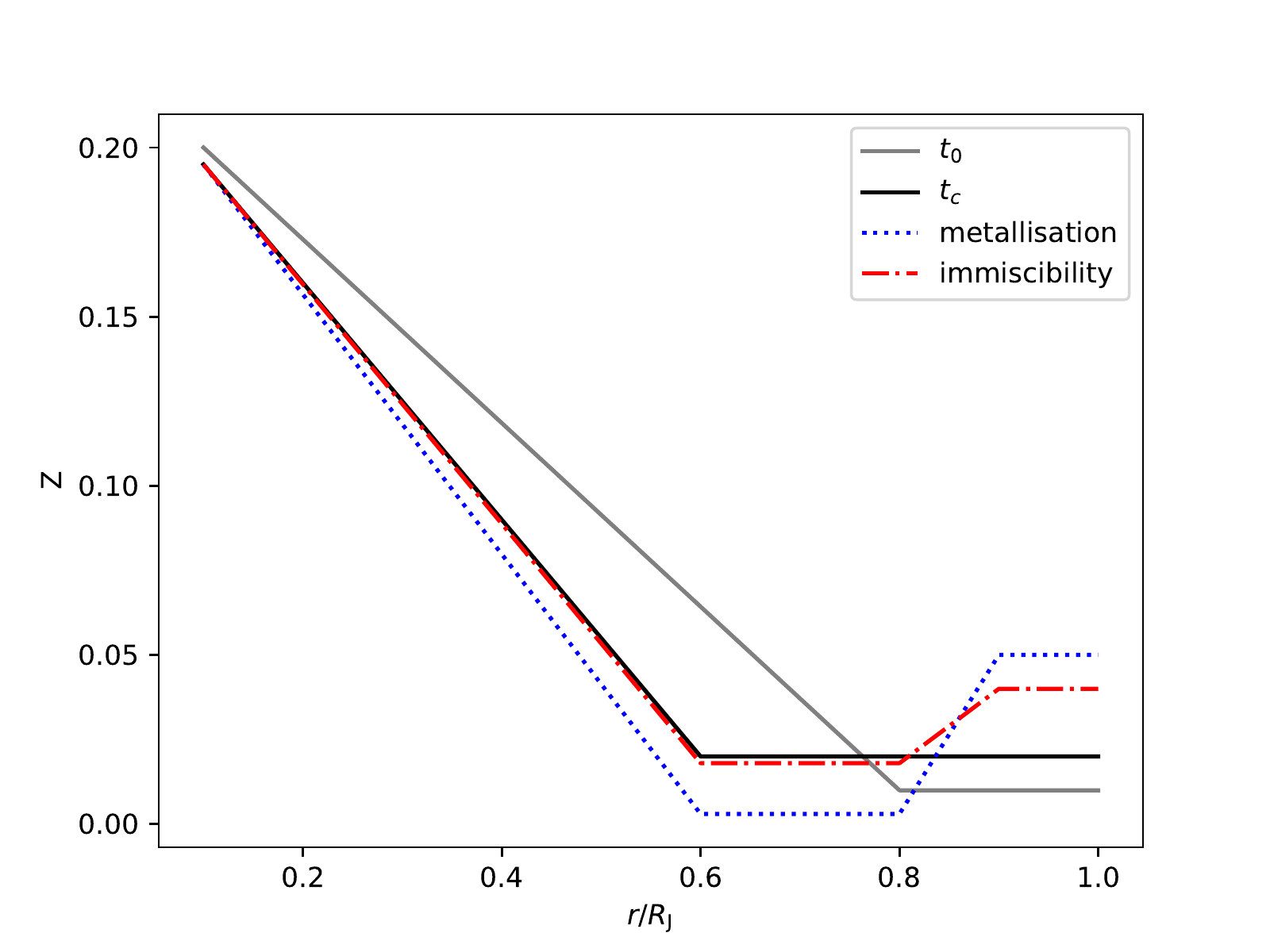}
\caption{Typical evolution of the heavy element content $Z$ as a function of the radial distance $r$
with time, assuming that a first order phase transition (PPT) or immiscibility
will occur during the cooling history. At $t = t_0$ Jupiter just formed. A small convective external envelope 
is connected to a gradually diluted planet, structure inherited from the core accretion. At $t = t_c$, 
immiscibility or first order metallisation is about to happen. The convective zone has somewhat expanded, 
redistributing the metal content in the planet. Later on ('metallisation' curve), 
the rising bubbles have enriched the outer envelope in heavy elements while depleted the inner one. If 
immiscibility takes place ('immiscibility' curve), the outer envelope will eventually lose part of its mass 
because of drowning helium dropplets, increasing the heavy elements mass fraction
in the outer envelope whereas the dilution of these elements in the inner envelope is almost negligible. 
\label{fig:evolution}}
\end{figure}

To be considered as plausible, our models with a region of locally inward decreasing abundance of heavy elements must 
be consistent with what is known of Jupiter's long term evolution. If, as expected, Jupiter formed through core accretion \citep{Pollack1996}, the primordial 
abundance of heavy elements in the planet should be increasing with depth (see Fig.\ref{fig:evolution}). As explored by, e.g., \citet{Leconte2013}, \citet{Vazan2018}, the differential core-envelope cooling 
of the planet leads to a redistribution of heavy elements with time, yielding an 
increasing heavy element content in the gas rich envelope. To explain our 
$\nabla Z >0$, a physical process must have inhibited convection within the envelope and 
prevented a homogeneous redistribution of elements. Three possibilities have been 
discussed in the previous subsections: 1$^{st}$ order metallisation, immiscibility and accretion.
We examine whether they are compatible with the evolution of the planet. 

If a first order metallization (PPT) stopped the convective motions, our discussion on the bubbles 
in \S\ref{sssec:1st_order_met} shows that it is possible to deplete the inner envelope and enrich
the outer one. This would be in adequation with any evolutionary scenario. 

In the case of hydrogen/helium immiscibility,
the outer envelope will be depleted in helium because of helium sedimentation (see \S\ref{thermo}, \citet{Stevenson1977}), yielding an enrichment in $Z=M_Z/M$ in this region.
Note that, in order to obtain a locally steep enough gradient of heavy elements $\nabla Z>0$
between the outer, $Z$-enriched, and the inner, $Z$-depleted, envelopes,
the He-rich falling dropplets must be as $Z$-poor as possible. This implies limited miscibility between 
some Z-components and He, in addition to H and He, and typical Z/He phase diagrams yielding very low concentations of heavy elements in He-rich dropplets, as discussed in \S4.1.1. Under such conditions, it is possible to preserve (and even increase) a positive $Z$-gradient with time. 

Finally, in the absence of PPT or immiscibility, if inhibition of convection, leading to layered convection (\citet{Leconte2012}) or even partly radiative interiors during the planet's growth,
is the only reason for the  difference of composition between the outer and inner envelopes, 
it must have persisted since the accretion event(s). 
Although, as examined in \S\ref{layer}, the present models fulfill the constraints required for the onset and persistence of layered convection (\citet{Leconte2012}, \citet{Leconte2013}),
whether such structures can persist during Jupiter's, or in fact any gaseous planet cooling history \citep{Chabrier2007} needs
to be explored with extreme care (e.g. \citet{Rosenblum2011}, \citet{Mirouh2012}, \citet{Wood2013}, \citet{Kurokawa2015}) and
requires that the key physical processes at play are handled with great accuracy. A fantastic challenge for numerical simulations.

To conclude this section, we should mention that recent evolutionary calculations (e.g. \citet{Vazan2018}) converge to a structure profile for present Jupiter with a monotonically
outward decreasing compositional gradient, $\nabla Z<0$, at odd with our suggestion of a local $\nabla Z>0$. These models, however, use 
the SCvH EOS. As discussed in \S\ref{entropy}, this latter yields a too hot thermal structure along an adiabat (see \citet{MH13} or Fig. 27 of \citet{Chabrier2018}), favoring convection and allowing larger
metal fractions. These Jupiter internal structure models are excluded by the present Juno+Galileo analysis and thus can not be used as reliable evolutionary constraints. Further
evolutionary calculations with the proper physics, including a proper treatment of double diffusive convection and of H/He phase
separation are needed to verify the consistency of the present models with Jupiter's thermal history. As just mentioned, however, properly handling
such complex physical processes is a task of major difficulty (see \S\ref{evol2}).

\section{Models with at least 4 layers and an entropy discontinuity 
in the gaseous envelope}
\label{sec:final}
\label{ssec:entropy}

As shown in \S\ref{sec:simple}, Jupiter interior structure must entail at least four different regions, namely two outer and inner homogeneous adiabatic envelopes,
separated by a region of compositional, thus entropy variation, and a diluted core, also harboring a more or less extended domain of compositional/entropy change, and potentially a solid rocky seed. One of the unknowns in these models is the amount of entropy change in the envelope. With an entropy change only due to a change in composition (see section \ref{sec:method}), 
the smallest value of $J_4$ we could obtain lies within the limit 
of what can be explained by a differential rotation shallower than
$10 000$ km \citep{Kaspi2017}. The higher
order moments, however, remain much too large. This yields another conclusion:

\begin{itemize}
\item Models with 
a {\it small} entropy change in the Mbar region seem to be excluded as possible Jupiter internal structure.
\end{itemize}

Therefore, in this section 
the inward increase of entropy (due to H-He immiscibility or the onset of super adiabatic layered convection)  is now used as a free parameter in the calculations, and is discretised throughout 
a certain number of spheroids across the ionization boundary region. That is we assume an entropy gradient
$\nabla S=\Delta S/\Delta R<0$ within the relevant pressure range. 

\subsection{Physical expectations}
\label{ssec:entropy_physics}

As quickly examined in section \ref{ssec:composition_change}, a brutal inward increase of $S$ can have several physical foundations. Assuming that Jupiter's outermost thermal {}profile is isentropic (because of adiabatic convection in this region), the observed condition, $T=165$ K, $P=1$ bar corresponds, according to our EOS, to $T\simeq 5000$ K at 1 Mbar.
As mentioned above, recent 1$^{st}$-principle simulations \citep{Mazzola2018} predict 
a critical temperature for the metallisation of hydrogen in the range $T_c\simeq 2000$-5000 K at $P\simeq$ 1 Mbar.
Both simulations (\citet{Soubiran2013}, \citet{Mazzola2018}) and experiments \citep{Loubeyre1985}, however, suggest
that, even for a  low helium concentration as in Jupiter ($x_{He}<0.1$), the critical pressure increases while the critical temperature
decreases with increasing helium concentration, which probably excludes a PPT between
molecular and metallic hydrogen in Jupiter. However, given
the present uncertainties in these determinations, we must still explore such a possibility.

In case of a 1st order transition, $\Delta S$ is given by the Clausius-Clapeyron relation along the critical line $P(T)$:

\begin{equation}
\frac{\Delta S}{\Delta (1/\rho)} = \frac{\mathrm{d} P}{\mathrm{d} T}.
\end{equation}
Analytical calculations  \citep{Saumon1992}, suggest 

\begin{equation}
\Delta S_\mathrm{metallization} \sim 0.5\, \mathrm{k_{\mathrm{B}}/proton}.
\end{equation} 
Since, as mentioned above, 
the temperature in this region of Jupiter's interior should be close to $T_c$, 
we expect $\Delta S$ to be less than this value. 

If hydrogen pressure ionization does not occur through a first order transition inside Jupiter,
a sharp entropy change
can be due to H/He phase separation (also a 1$^{st}$-order transition). As shown by \citet{Stevenson1977}, drowning nucleated helium dropplets
lead to a release of gravitational 
energy and, even though their analysis suggest that most 
of this energy is radiated away, part of it contributes to heating up the inner part of the planet, raising the entropy
(see detailed discussion in \S\ref{evol2}).




The shape of the H/He phase diagram is a major uncertainty in this context.
The rather limited helium depletion w.r.t. the solar value in Jupiter's external envelope, $x_{He}\simeq 0.1$, suggests that
the variation $\Delta Y$ in the immiscible region should be modest (about $\sim 10\%$).
In that case, according to \citet{Morales2013} Fig. 2, the mixing entropy should depart
only slightly from the ideal mixing entropy, by $\sim 0.03\,k_B$/at at 5000 K for $x_{He}=0.1$. 
Since the maximum value of the
ideal mixing entropy, for a concentration $x_{He}=0.5$ is $S_{mix}^{id}/N=-[x\ln x+(1-x)\ln (1-x)]=0.7\,k_B$/at (about $0.3\,k_B$/at for $x_{He}=0.1$), we see that the entropy jump
due to H/He immiscibility should be $\lesssim 0.5\,k_B$/proton.
The entropy change due to helium dropplet sedimentation is more difficult to evaluate and requires numerical explorations.
Guidance is provided by the calculations of \citet{Fortney2003} for the case of Saturn. In the case of a maximum temperature gradient in the
inhomogeneous region and no formation of a helium layer atop the core (both the most likely present situation), these authors find that
a change of composition $Y=0.21 \rightarrow 0.36$ corresponds to a global increase of entropy $\Delta S\sim 0.3\,k_B$/proton. For Jupiter, we expect helium sedimentation (i) to have occured,
if ever, more recently than for Saturn, (ii) to encompass a much smaller fraction of the planet (see \S\ref{HHe}) and thus to induce a much smaller entropy variation. 
Adding up these two contributions, it seems difficult to justify an entropy jump arising from H/He phase separation much larger than:

\begin{equation}
\Delta S_\mathrm{H/He} \lesssim 1.0\, \mathrm{k_{\mathrm{B}}/proton}.
\end{equation} 
Clearly, more experimental and numerical exploration of hydrogen pressure metallisation and H/He phase diagram and He sedimentation process are strongly needed to help constraining these processes.

Finally, if  the 
mean molecular weight gradient due to the change of composition is large enough to hamper adiabatic convection, 
a regime of layered convection can develop and lead 
to a super adiabatic temperature structure similar, at least in some part of the planet, to the one derived in \citet{Leconte2012}. 
The detailed treatment of layered convection in our calculations is presented in \S\ref{layer}.
Varying the 
location and extent of layered convection between $0.1$ Mbar and $2$ Mbar, i.e. in the vicinity of hydrogen pressure ionization, we obtain numerically  
a maximum entropy increase from layered convection:

\begin{equation}
\Delta S_\mathrm{layered} \lesssim 0.6\, \mathrm{k_B/proton},
\end{equation}
with a decreasing metal abundance, i.e. $\nabla Z>0$, in this region. An 
increasing metal abundance in this region yields higher values of $\Delta S$ but in that case $\alpha \lesssim 10^{-7}$, where $\alpha$ denotes the ratio of the size of the convective layer
to the pressure scale height, $\alpha=l/H_P$ (see \S\ref{sec:discussion}). This implies 
the presence of a diffusive buffer, or a regime of turbulent diffusion within the ionization boundary layer. Although detailed calculations are lacking, it seems
difficult to reconcile such a structure with Jupiter's thermal history. 

In summary, if either a first order transition and/or layered convection is present within some part (most likely around the Mbar) of the planet gas rich envelope, an
inhomogeneous zone where the total increase in entropy can reach  $\sim 0.5$-$1\, \mathrm{k_B/proton}$ is expected.

\begin{figure*}[ht!]
\begin{subfigure}{.5\linewidth}
  \flushleft
  \includegraphics[width=\linewidth]{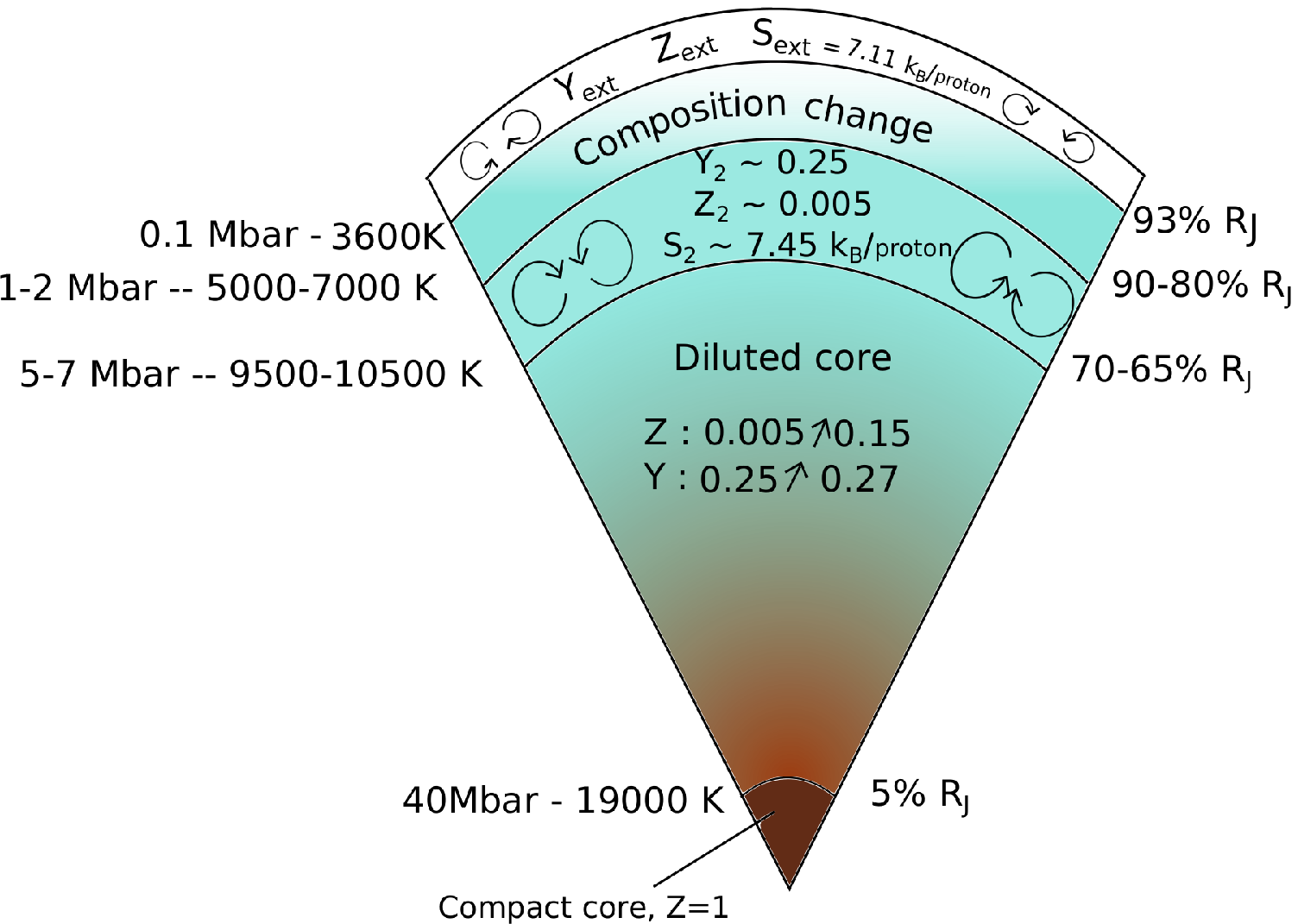}
  \caption{}
  \label{fig:final_low}
\end{subfigure} 
\begin{subfigure}{.5\linewidth}
  \flushright
  \includegraphics[width=\linewidth]{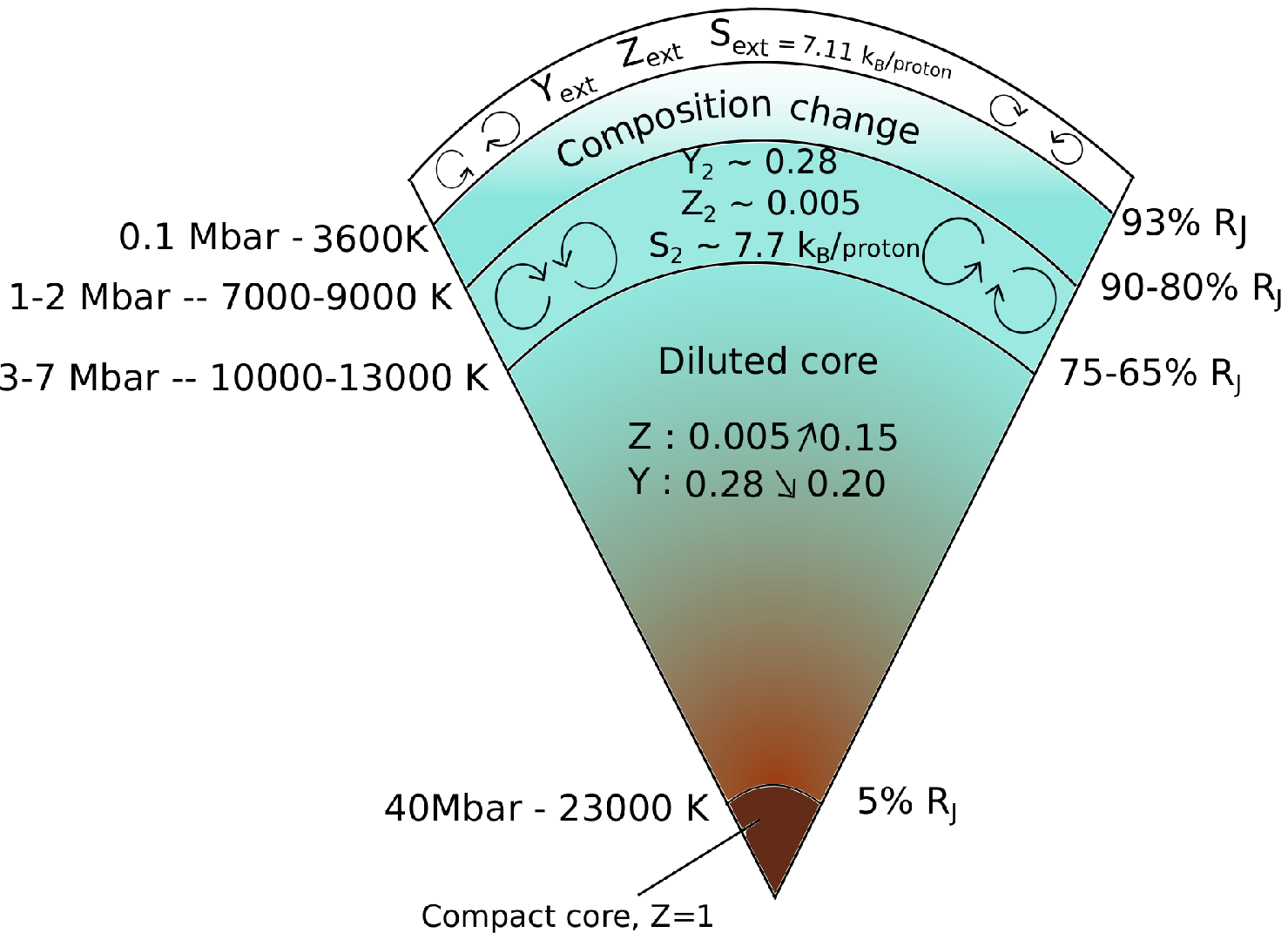}
  \caption{}
  \label{fig:final_normn}
\end{subfigure}
\\
\begin{subfigure}{\linewidth}
  \centering
  \includegraphics[width=.5\linewidth]{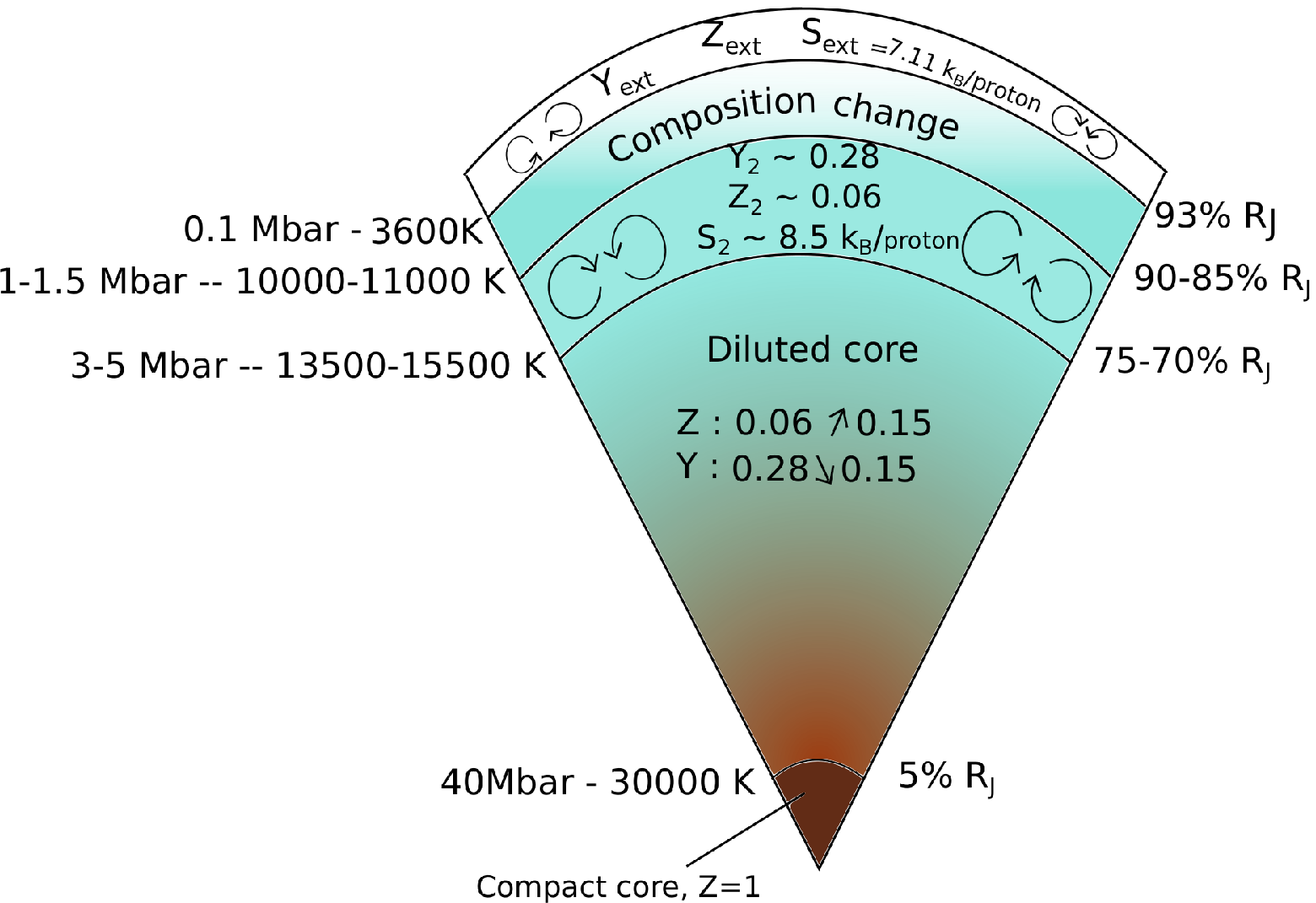}
  \caption{}
  \label{fig:final_grad}
\end{subfigure}
\caption{Schematic internal structure of our final Jupiter models. $Y_\mathrm{ext} = 0.23$, $Z_\mathrm{ext} = 0.02$
as stated in the text. (a) The modest entropy increase 
between the outer and inner envelopes yields a moderate helium 
increase in this latter, and an inward helium enrichment in the diluted core (see text). (b) The larger  entropy increase in
the inhomogeneous boudary region yields a supersolar helium
fraction in the inner envelope, but then the helium abundance decreases in the diluted core. (c) Our least
favoured model. An increase of both helium and heavy element abundances in the inner envelope requires
a strong entropy increase, at the limit of what is physically achievable. A mixture of 
structures (a) and (c) is also possible, with a small increase in both helium and heavy 
elements. The required $\Delta S$ would be comparable to (b).}
\label{fig:final}
\end{figure*}

Figure \ref{fig:final} portrays
the typical structure of our final Jupiter models. They all share the following features:
\begin{itemize}
\item an outer homogeneous convective envelope characterised by the Galileo helium and heavy element abundances and the adiabat condition at $T=165$ K, $P=1$ bar;
\item an inhomogeneous region between $\sim$0.1 and $\sim$2 Mbar associated with (i) a change in composition, most likely characterised by an {\it inward decreasing metal abundance} ($\nabla Z>0$) and (ii) 
a non negligible entropy ($\gtrsim 5\%$)  and temperature increase. These gradients stem
from layered convection and/or H-He immiscibility, even though a PPT can not be totally excluded for now;
\item an inner homogeneous convective envelope lying on a warmer isentrope than the outer region, with an average larger helium fraction and,
most likely, a lower metal fraction than in the outermost region. Indeed, even though a larger $Z$ fraction in this region than in the outer one 
is not entirely excluded, it
requires an uncomfortably large entropy increase, $\Delta S\gtrsim 1.4\,k_B/$proton (case (c)), according to the aforederived estimates.

As shown below, there is a degeneracy between the entropy jump in the inhomogeneous region and the helium and metal fractions in the
inner envelope. The larger $Y$ and $Z$ in the inner envelope, 
the larger $\Delta S$ needs to be.
\item a diluted (eroded) core extending throughout a significant fraction of the planet. A small entropy jump in the inhomogeneous region,
$\Delta S\lesssim 0.5\,k_B/$proton (case (a)), yields an inward increasing helium abundance in the core, while a larger value (case (b)) implies an inward decreasing helium abundance in the core.
\item probably, but not necessarily, a central compact, solid core.
\end{itemize}

A quantitative analysis of these models is given in the next subsection.

\subsection{Quantitative results on the gravitational moments}

\begin{figure*}[ht!]
\begin{subfigure}{.5\linewidth}
  \flushleft
  \includegraphics[width=\linewidth]{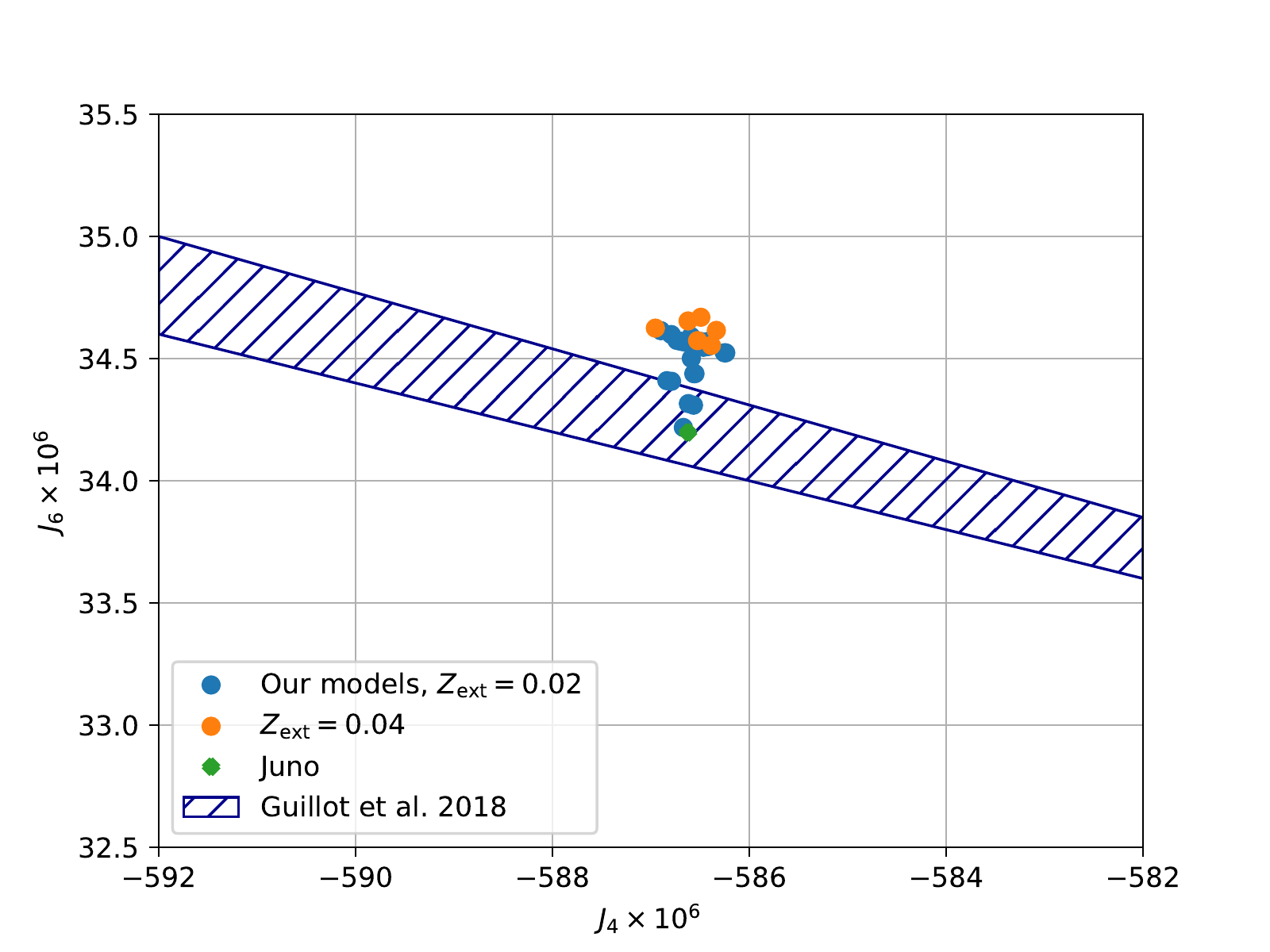}
  \caption{}
  \label{fig:J4J6}
\end{subfigure} 
\begin{subfigure}{.5\linewidth}
  \flushright
  \includegraphics[width=\linewidth]{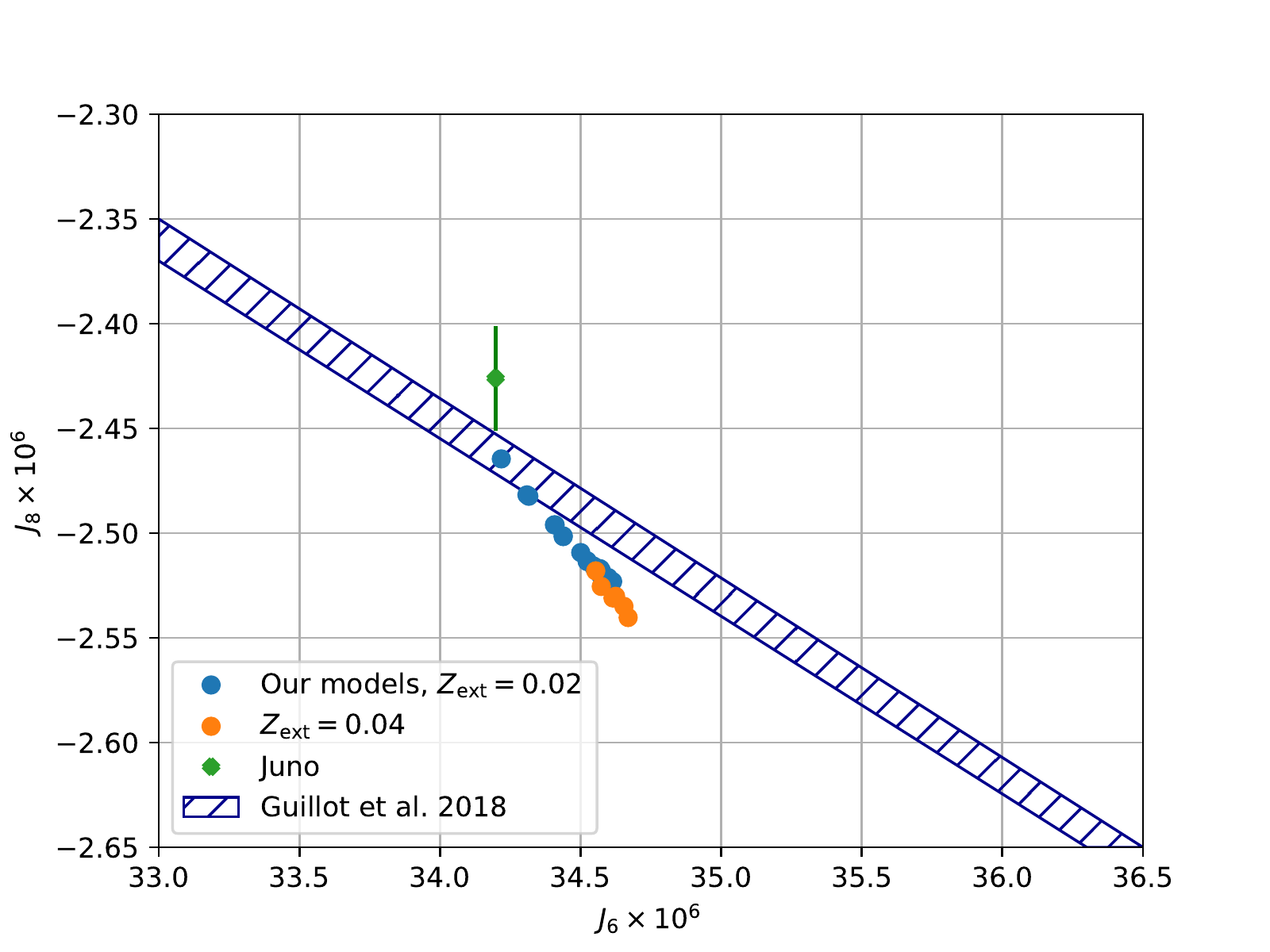}
  \caption{}
  \label{fig:J6J8}
\end{subfigure}
\\
\begin{subfigure}{\linewidth}
  \centering
  \includegraphics[width=.5\linewidth]{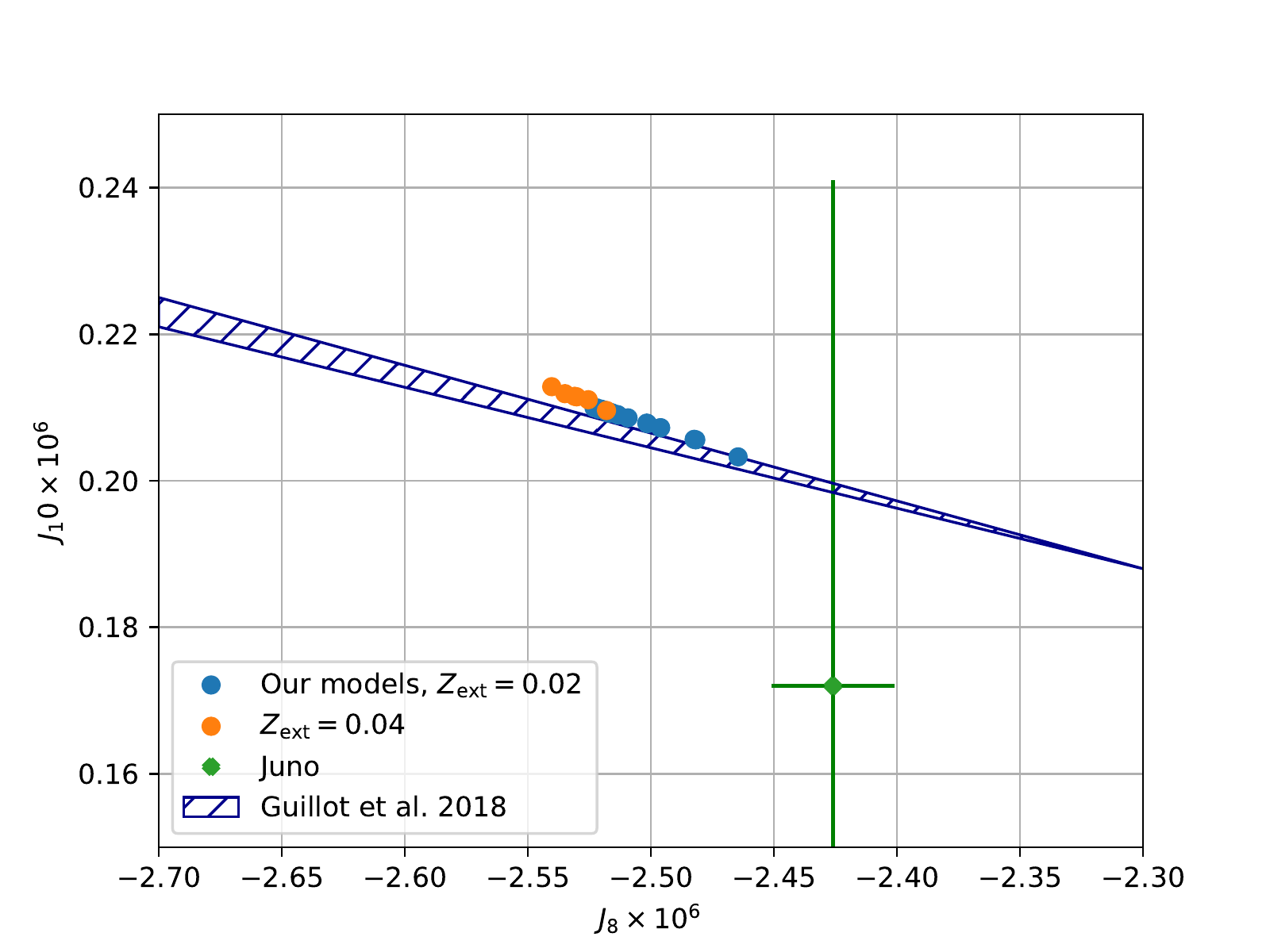}
  \caption{}
  \label{fig:J8J10}
\end{subfigure}
\caption{Gravitational moments we obtain with various models 
in the $J_k - J_{k+2}$ plans, for two values of $\zext$. All the values are multiplied by $1 \times 10^6$
as done in e.g., \citet{Guillot2018}. The green dot is the Juno value with the
observed error bars \citep{Iess2018}, without any dynamical correction. The hatched area 
corresponds to the models of \citet{Guillot2018}. \label{fig:JkJk2}}
\end{figure*}

The results of our optimized models with an entropy discontinuity $\Delta S \in [0-2] \mathrm{k_\mathrm{B}/proton}$ in the ionization boundary region, projected
in different $J_k$ - $J_{k+2}$ plans, are displayed in Figure \ref{fig:JkJk2} for two values of external heavy element abundance, namely
$\zext=Z_\odot$ and $\zext=2\times Z_\odot$.
The first obvious conclusion from this figure is that 
our range of models consistent with Juno's observed gravitational moments differs
from the one derived by \citet{Guillot2018} with 200 000 models
(see their Fig. 1 of the Extended Data, reported as a hashed area in Fig. \ref{fig:JkJk2}). 
We have verified that this is not a discretisation issue: with interior structures
calculated with 1000 spheroids, this conclusion is unaltered.
Even though the difference between the two analysis should partly stem from
the different EOS used by these authors, it arises essentially from our different representation of the planet interior. 
Indeed, we recall that these authors do not take into account the constraint from Galileo on the heavy element abundances. 
Therefore, if Galileo's observations are correct, Fig. \ref{fig:JkJk2} shows that 
the qualitative conclusions these authors draw about differential rotation could be altered.

\begin{figure*}[ht!]
\begin{subfigure}{.5\linewidth}
  \flushleft
  \includegraphics[width=\linewidth]{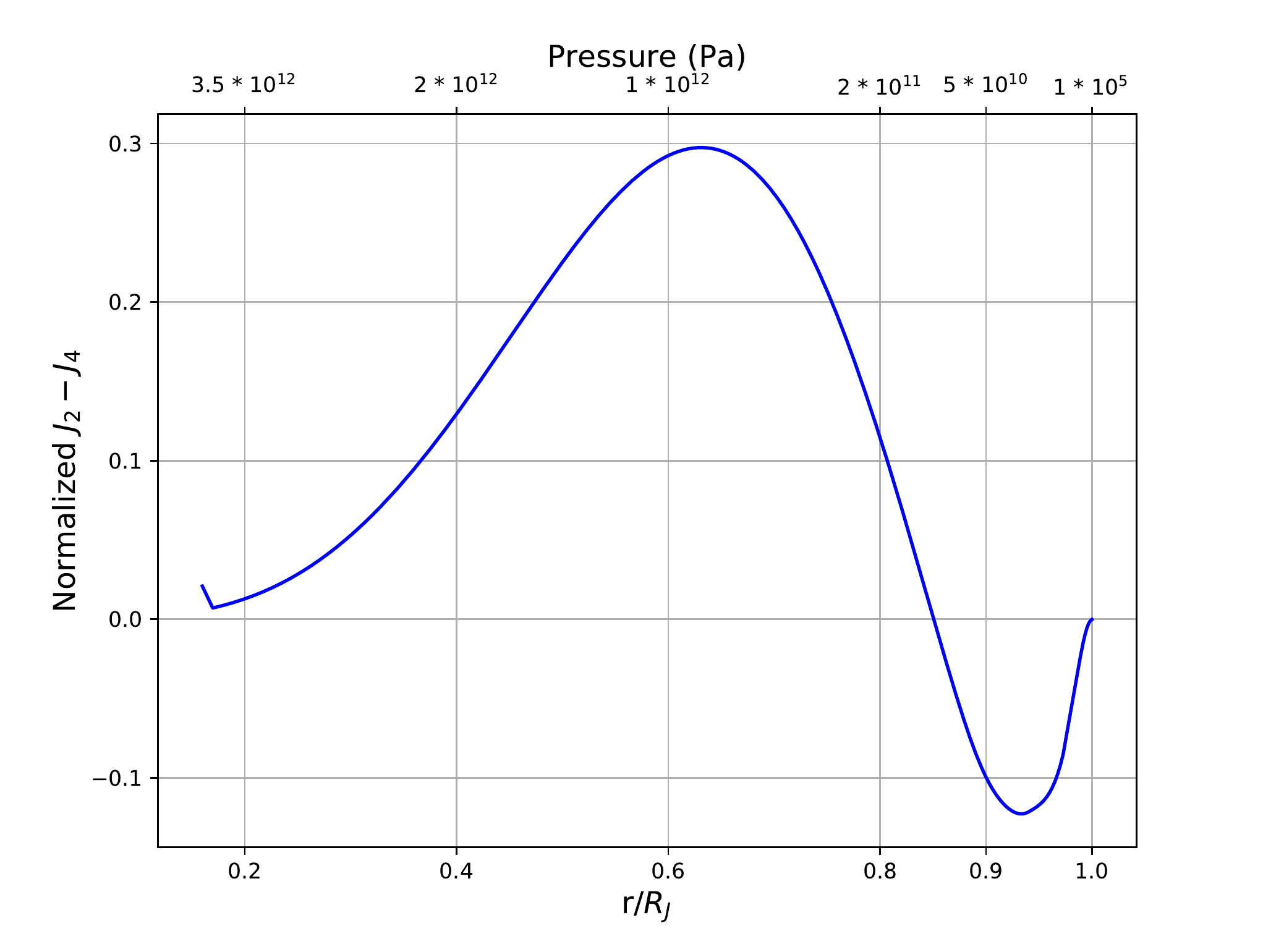}
  \caption{}
\end{subfigure} 
\begin{subfigure}{.5\linewidth}
  \flushright
  \includegraphics[width=\linewidth]{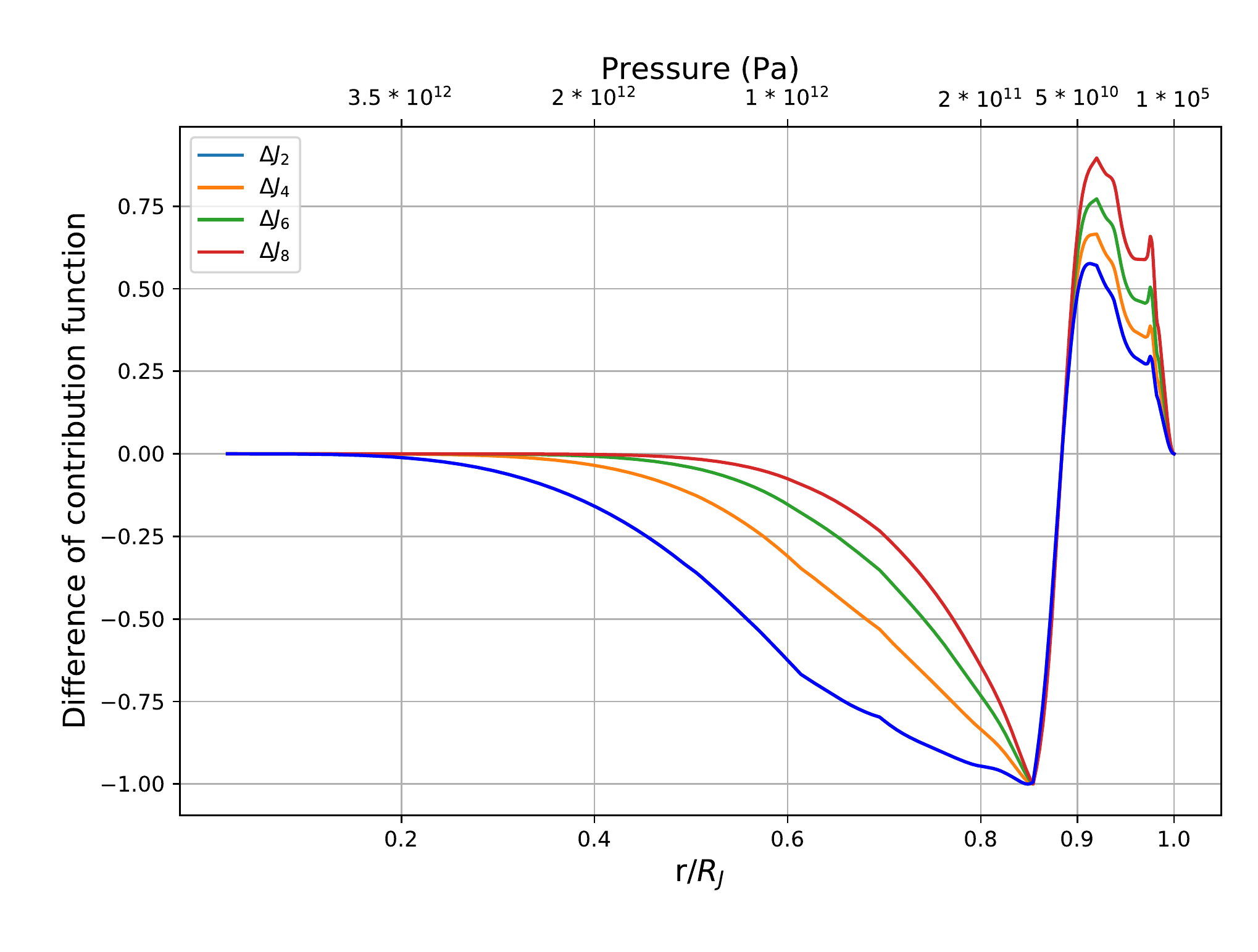}
  \caption{}
\end{subfigure}
\caption{(a) Subtraction of the contribution functions
 for $J_2$ minus $J_4$, both normalized to their maxima, with respect 
to pressure or radius, for an isentropic model. (b) Normalized difference of contribution functions for $J_2$ to $J_8$ between
a model with a small entropy change and a model with a strong entropy change.}
\label{fig:J2-J4}
\label{fig:contributions}
\end{figure*}

- {\it Impact on the low-order moments ($\le J_4$)}. We found out that the entropy change, $\Delta S$, is strongly affected 
by the composition in the inner convective envelope, i.e.  the region between the inhomogeneous one and the diluted core. 
Therefore the size and composition of the diluted core, the composition
of the inner envelope, and the entropy change $\Delta S$ in the region of compositional variation are intrinsically linked.
To better understand this
result, we must recall
that the major issue of the models is to decrease $J_4$ at constant $J_2$. Figure \ref{fig:J2-J4}(a) portrays
the value of the contribution function $J_2 - J_4$ in the planet
as a function of pressure.

This figure shows that, in order to decrease $J_4$ with respect to $J_2$, one
needs to enrich the planet deeper than $\sim 2 \,\mathrm{Mbar}$, and the region around $\sim 10 \,\mathrm{Mbar}$
is where it is most efficient. 
Therefore, an enriched inner envelope decreases $J_4$ at constant 
$J_2$ (see the pressure range in Figure \ref{fig:J2-J4}(a)); but enriching the inner envelope
implies a steeper compositional gradient in the boundary region between the outer and inner envelopes, which has the opposite effect 
on $J_4$ compared to $J_2$. Furthermore, this boundary region between $\sim$0.1 and 
2 Mbar has a much stronger contribution on $J_2$ and $J_4$ than the deeper region. This stems from the fact that this region has 
a high mean radius, hence the mass of a spherical shell is much 
larger than in the 5 Mbar region, and the impact on $J_2$ and $J_4$ is enhanced. 
In consequence, a small change in the $\sim 0.1-1$ Mbar region must be compensated
by a strong change in the diluted core.

- {\it Impact on the high-order moments ($> J_4$)}. The high order gravitational moments strongly depend on the value of $J_4$, as shown by \citet{Guillot2018}. 
For a given $J_4$, the other parameters affecting these moments are the 
external abundance of metals (as expected), and the mass of the central 
compact core. 
Changing the helium content within the inner convective envelope has almost no impact, as there is a trade-off
between the inner abundances of helium and heavy elements and the 
entropy increase, without affecting the high order
gravitational moments. Similarly, the position and extent of the boundary region of compositional change
is a second order correction to the $J_6$ to $J_{10}$ values.
As a whole, we found out that the $J_6$ to $J_{10}$ values are not 
much affected by the composition in the inner part of Jupiter, deeper than where the compositional change occurs.

As mentioned above, we found out that  some
models with an {\it inward increase} of heavy elements in the envelope inhomogeneous region ($\nabla Z<0$) can fulfill all observational constraints (case (c)) 
provided the entropy change around $\sim 1$ Mbar reaches values $\Delta S>1 k_\mathrm{B}/\mathrm{proton}$.
This requires a strong entropy discontinuity induced either by a PPT for $T\simeq 4500$ K (thus a critical point $T_c\gg 4500$ K) or by H/He differentiation and sedimentation, as layered convection alone cannot yield such an entropy jump. Therefore, {\it although not entirely excluded}, models with $\nabla Z<0$ throughout the entire envelope are rather uncomfortable, as discussed in \S\ref{ssec:entropy_physics}. In contrast, models  with an {\it inward decreasing} abundance of heavy elements  ($\nabla Z>0$) in this region  require a more modest entropy change. 

The fact that, surprisingly, the mass of the {\it compact} core affects the high order 
moments can be explained as follows. Since we consider the central compact core 
as spherical, it has no direct influence on the gravitational moments.
However, in that case, a smaller fraction of the planet's mass is available to satisfy the $J$ values. Since the outer envelope composition is constrained
by Galileo, one can only enrich the inner envelope or the {\it diluted}
core to compensate. Fig. \ref{fig:J2-J4}(a), however, shows that 
if there is a too large increase of density deeper than $2$ Mbar,
the increase of $J_2$ is larger than the one of $J_4$ (and even larger
than the increase of the higher order moments, not shown). This leads to  
\\

$\bullet$ {\it Conclusion 1}:
{\it for given $J_2$, a central compact core tends to decrease the moments of order $\ge J_4$ compared to a model with only a diluted core}.

Increasing the mass of the compact core thus implies 
to add more heavy elements in the inner regions of the planets, diluted core or inner envelope to reproduce the $J$ values.
But the larger the amount of heavy elements
in the deep layers the larger the required entropy jump $\Delta S$ between the outer and inner envelopes.
This, in turn, has consequences on the high order moments:  
a model with a high $\Delta S$ in the metallisation region implies a
higher temperature, and thus a lower density for a given composition
at given pressure in the Mbar region than a model with a smaller $\Delta S$. The lower density means that this region
has a smaller contribution to the gravitational moments than in the case of a small $\Delta S$.
Although, for $J_2$ and $J_4$, this can be balanced
by a higher contribution of the internal layers (deeper than a few Mbars), 
this is not the case for the higher order moments.

To illustrate this result, we show in Figure \ref{fig:contributions}(b) the differences $\Delta J_n$ in
the contribution functions $J_2$ to $J_8$ 
between a model with a small entropy change, $\delta S<0.1\,\mathrm{k_\mathrm{B}/\mathrm{proton}}$, in the ionization boundary region,
only due to a composition change, and a model
with a total entropy discontinuity $\Delta S=0.9\, \mathrm{k_\mathrm{B}/\mathrm{proton}}$. 
We see that the region of the outer (molecular) envelope has always a stronger contribution
to the $J$'s
when the entropy change is small, as expected, whatever the order of the moment. On the other hand, the inner region of the diluted core
has much more impact on $J_2$ than on the other moments. 
A strong $\Delta S$ thus requires more heavy elements 
in the diluted core to preserve $J_2$ while the high order moments are almost insensitive
to the enhanced composition in the diluted core. This leads to  

$\bullet$ {\it Conclusion 2}: {\it 
an entropy jump in the envelope tends to decrease the value of the high order 
moments at a given $J_2$.} And a large enough entropy change is necessary to preserve the correct balance between the moments.
\\

\begin{figure*}[ht!]
\begin{subfigure}{\linewidth}
  \centering
  \includegraphics[width=.5\linewidth]{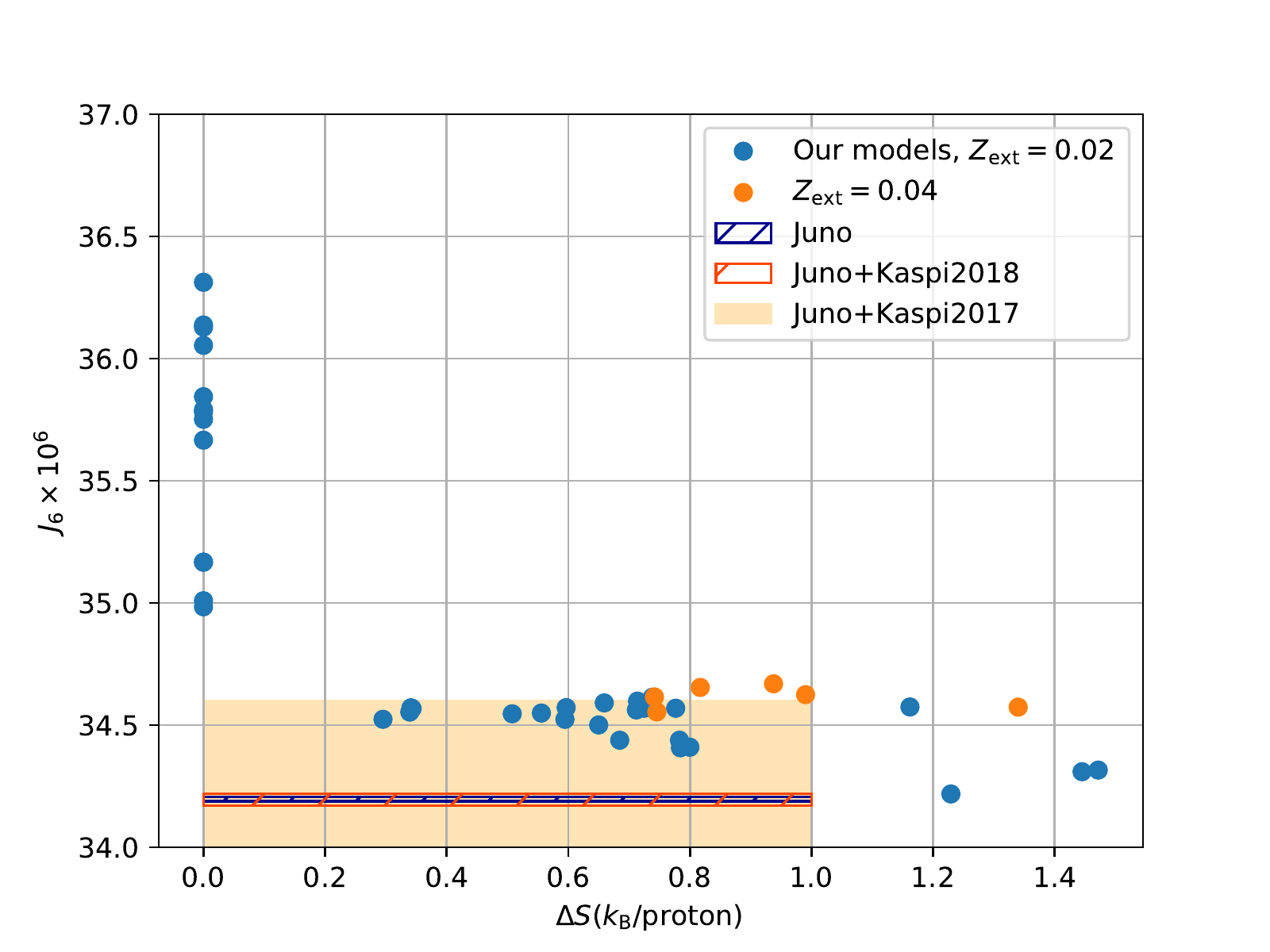}
  \caption{}
  \label{fig:dSJ6}
\end{subfigure} 
\\
\begin{subfigure}{.5\linewidth}
  \flushright
  \includegraphics[width=\linewidth]{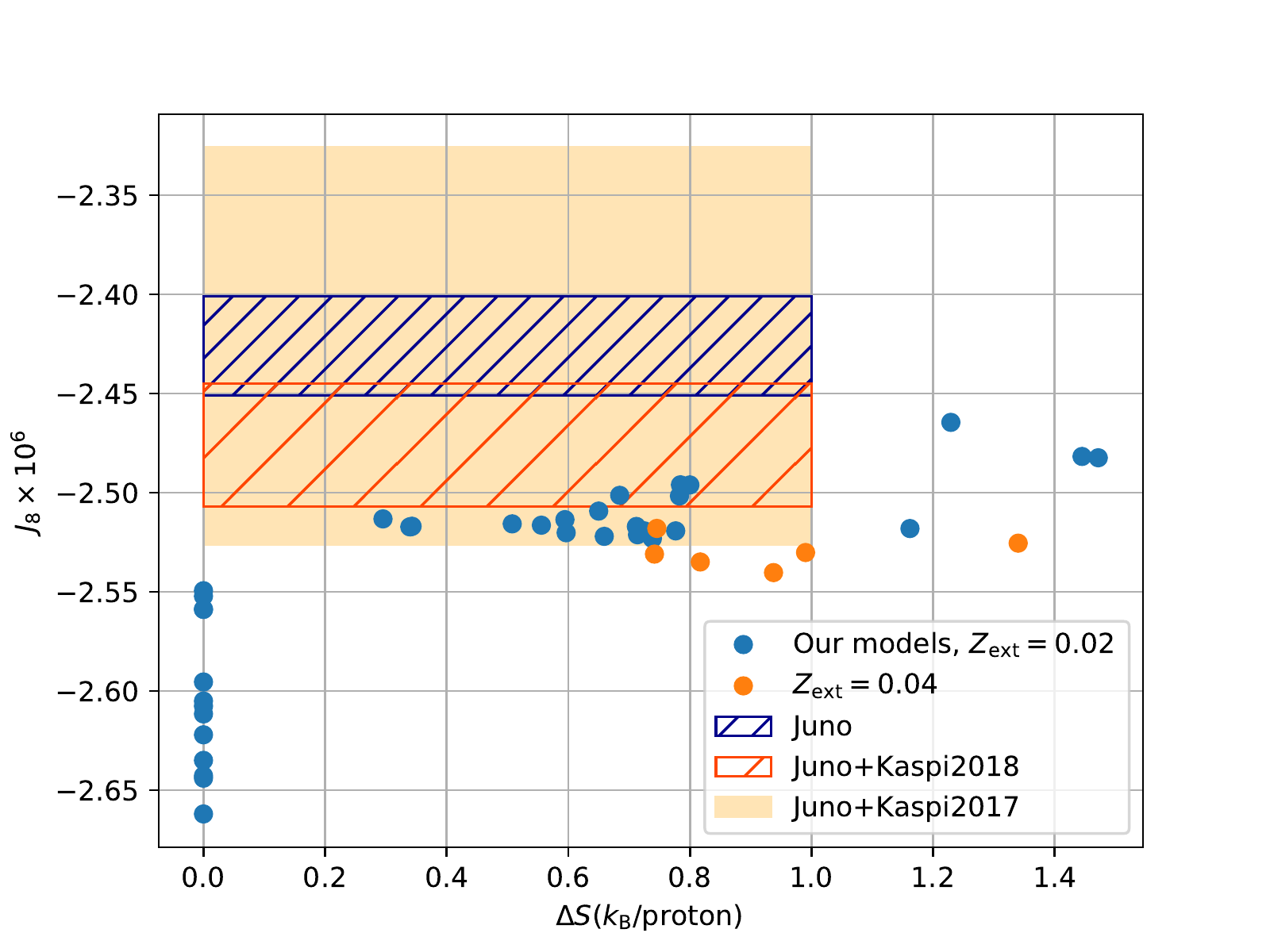}
  \caption{}
  \label{fig:dSJ8}
\end{subfigure}
\begin{subfigure}{.5\linewidth}
  \flushright
  \includegraphics[width=\linewidth]{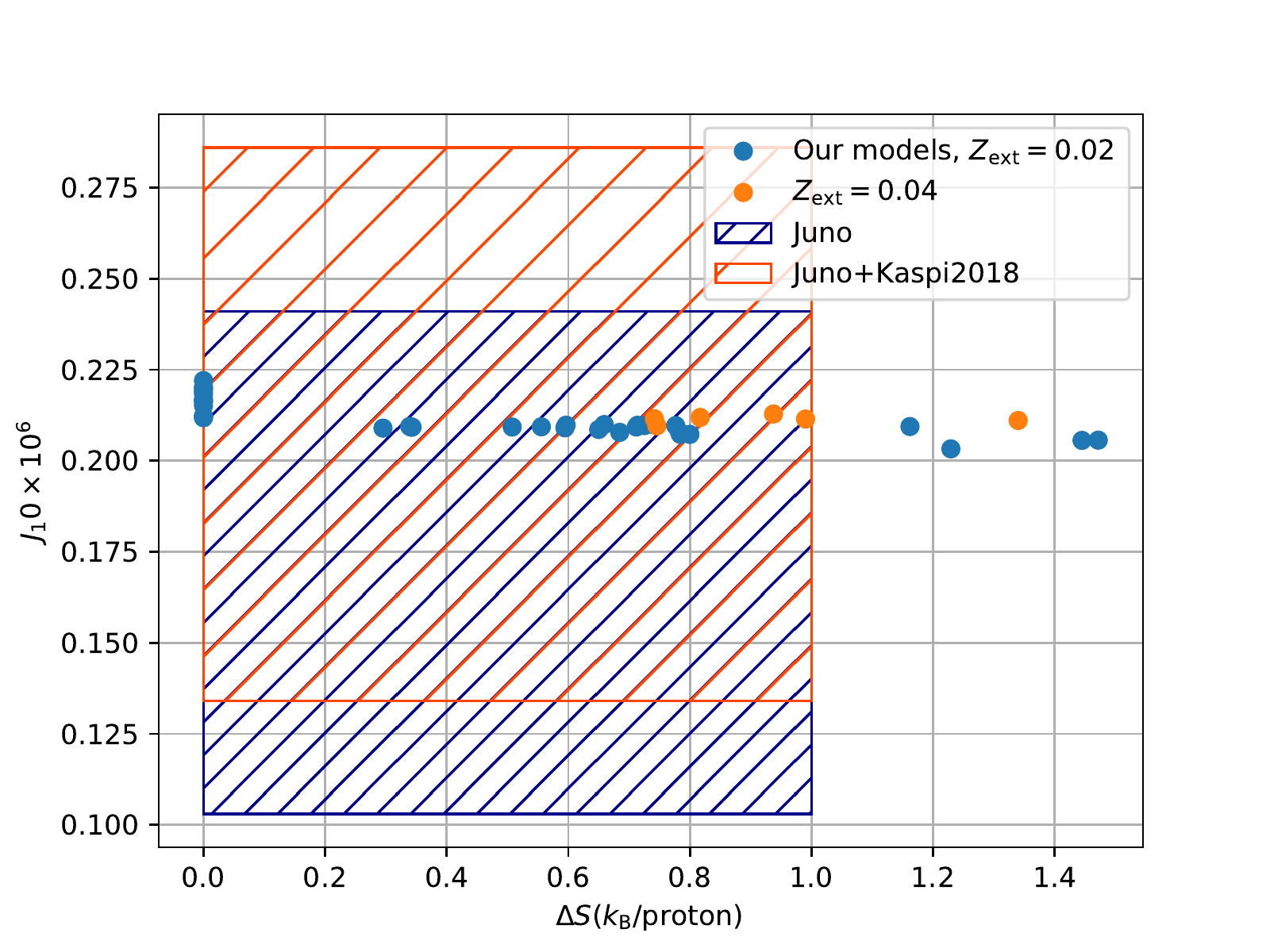}
  \caption{}
  \label{fig:dSJ10}
\end{subfigure}
\caption{High order gravitational moments as 
a function of the entropy jump $\Delta S$ in the envelope. All these models 
have $J_2$ and $J_4$ values within the allowed range from Juno's observations 
with the \citet{Kaspi2018} correction , except 
those with $\Delta S = 0$ for which we could not even match $J_4$.  The $Z_{ext}$ value is the atmospheric $Z$,
'Juno' corresponds to the observations of Juno with the $3 \sigma$ error bars, 
'Juno + Kaspi2018' are the observations corrected by the winds 
estimated in \citet{Kaspi2018}, and 
'Juno + Kaspi2017' includes the correction to the observed gravitational moments due 
to a differential rotation shallower than 3000 km, from \citet{Kaspi2017}. While several models are compatible with Juno's observations when taking into account
the corrections due to differential estimated by \citet{Kaspi2017}, this becomes much more difficult when considering
the correction derived from the odd gravitational moments by \citet{Kaspi2018}. Note that, in this latter case, 
none of the models with $\Delta S = 0$ or $\nabla Z <0$ can reproduce the gravitational moments of Jupiter.}
\label{fig:dS}
\end{figure*}

As discussed in \S\ref{ssec:entropy_physics}, however, the possible entropy increase in the metallization boundary region is limited by physics principles. 
For central compact cores larger than $ \gtrsim 5 \mearth$, one needs 
$\Delta S>1 \,\mathrm{k_\mathrm{B} /\mathrm{proton}}$, which,
as discussed above, seems to be hardly possible at these temperatures. Figure \ref{fig:dS} displays the values of the high order gravitational moments as a function of the entropy jump $\Delta S$. Small (absolute) values of $J_6$, $J_8$ or $J_{10}$ 
allways require a significant $\Delta S$, except 
if we decrease the atmospheric $\zext$, violating in that case Galileo's constraint, as done in all recent studies.
Models with no entropy jump in the gaseous envelope thus seem to be excluded, as mentioned previously.

As seen in Fig. \ref{fig:dS}, none of our 
models can match the $3 \sigma$ error bars on $J_6$ for $\Delta S<1 \,\mathrm{k_\mathrm{B}/proton}$ when considering the contribution 
from the winds derived in \citet{Kaspi2018}. This is particularly true if the external abundance
of heavy elements $Z_\mathrm{ext}$ is supersolar (see \S\ref{ssec:high_Z}). When considering the dynamical correction from \citet{Kaspi2017}, however, flows extending down to $3000$ km are sufficient to explain the discrepancy 
with the observed gravitational moments. Therefore, either the $\Delta J_6$ correction
due to the wind contribution  in \citet{Kaspi2018} is underestimated, 
because of an erroneous estimation of the winds or the presence of 
North-South symmetric zonal flows which will affect the even gravitational moments,
or the entropy 
increase must reach at least $\sim 1.5\,\mathrm{k_\mathrm{B}/proton}$. In any case, a continuously increasing heavy element
mass fraction with depth, i.e. $\nabla Z<0$, in the Mbar region is hard to justify (on Fig. \ref{fig:dS}, 
such models all have $\Delta S \gtrsim 1 \mathrm{k_\mathrm{B}/proton}$).

As shown in Fig. \ref{fig:final}, the valid models predict a size for the metallization boundary region, $l_b\approx 15\%$ of Jupiter's radius. Clearly,
this is orders of magnitude larger than any possible interface due to a PPT. It can, however, be consistent with the size of the inhomogeneous
H/He region, as this latter keeps expanding during the planet's cooling. Finally, as shown in
\S\ref{layer}, this region, characterised by a compositional gradient, is prone to layered convection, by itself characterised by a superadiabaticity and thus by
its own entropy variation, to be added up to the one issued from a phase transition, and thus contributing to the total $\Delta S$.

\subsection{Optimized Jupiter models}

\begin{figure*}[ht!]
\begin{subfigure}{.5\linewidth}
  \flushleft
  \includegraphics[width=\linewidth]{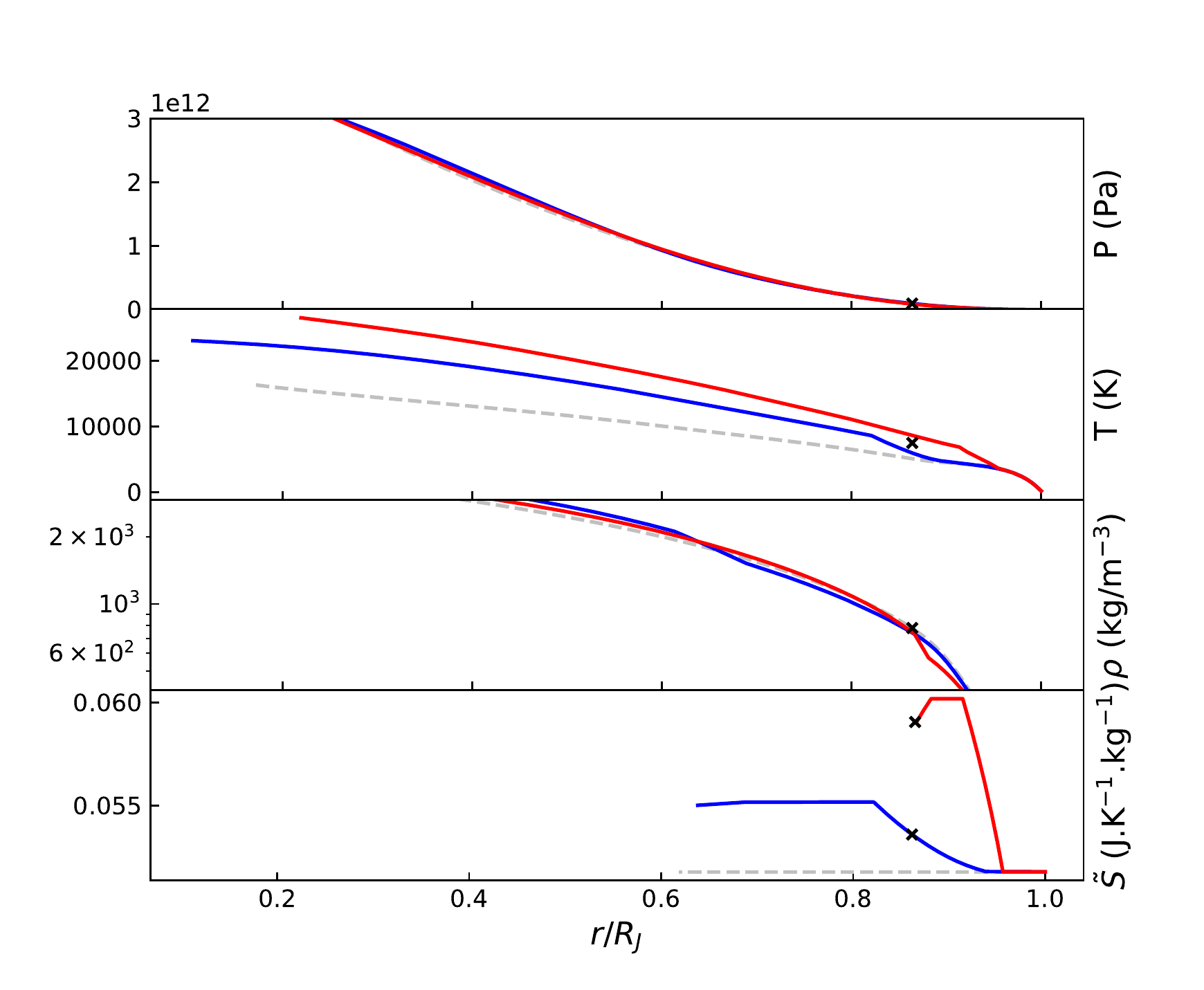}
  \caption{}
  \label{fig:profiles}
\end{subfigure} 
\begin{subfigure}{.5\linewidth}
  \flushright
  \includegraphics[width=\linewidth]{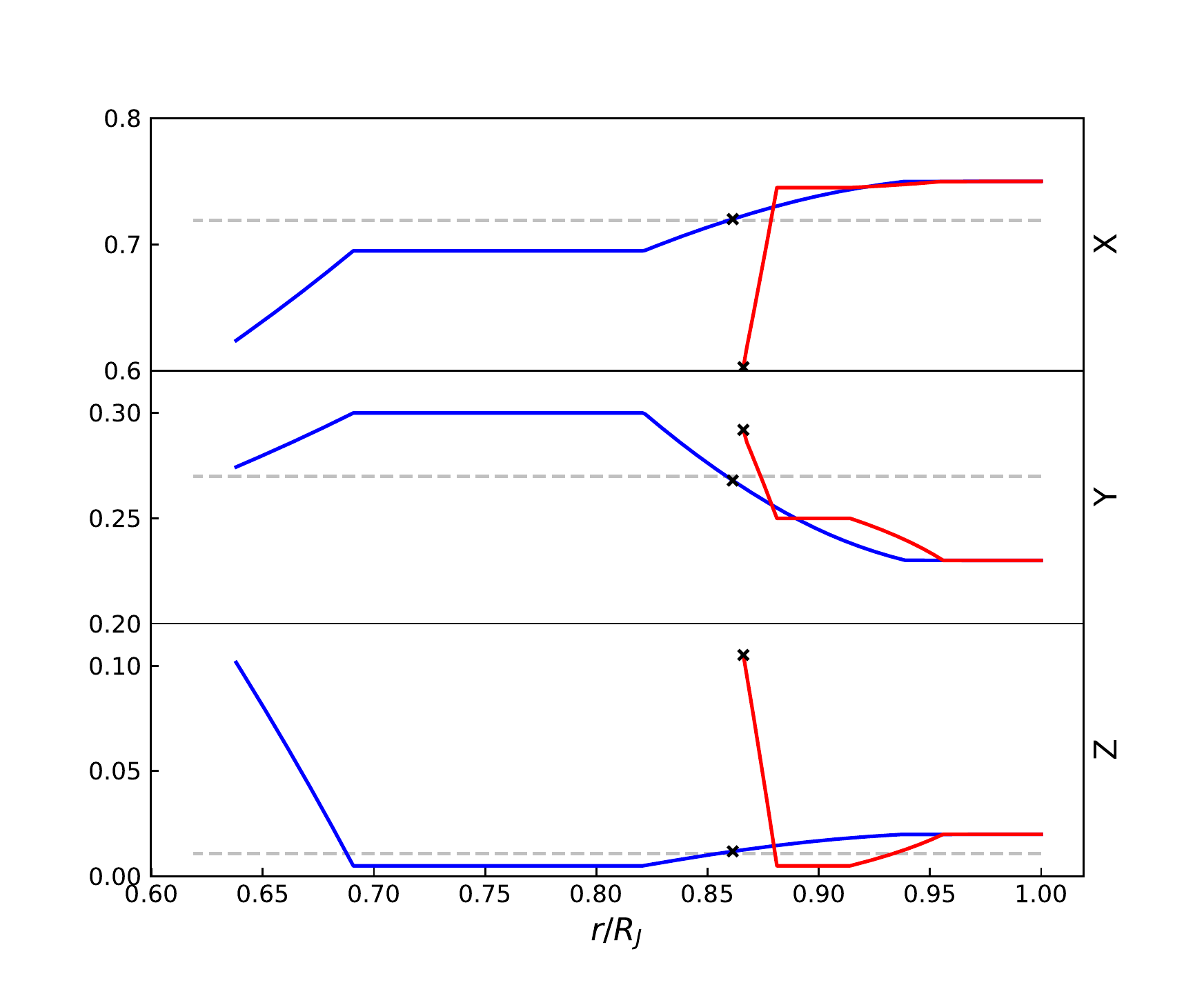}
  \caption{}
  \label{fig:composition}
\end{subfigure}
\caption{(a) Pressure, temperature, density and specific entropy as a function
of radius for an 
isentropic (dashed gray) structure of Jupiter, a model with a high $J_6$ value (blue), compatible with  \citet{Kaspi2017}, 
and a model with a lowered $J_6$ compatible with \citet{Kaspi2018} (red). (b) Hydrogen (X), helium  (Y) and heavy element (Z) mass abundances for the same models. The black crosses correspond
to $P=1 \mathrm{Mbar}$, about the region of hydrogen pressure dissociation/ionization. The outer and inner convective zones correspond to the regions of constant (homogeneous) composition and entropy, whereas the gradients are representative of the envelope inhomogeneous region and the outer part of the diluted core, respectively.}
\label{fig:models}
\end{figure*}

Figure \ref{fig:models} portrays the thermodynamic and composition profiles
of our models consistent with all Galileo and Juno constraints, taking into account for this latter the correction due to differential rotation from \citet{Kaspi2017} and \citet{Kaspi2018}, respectively. Profiles for an isentropic interior structure, inconsistent of course with the observed gravitational moments, are shown for comparison. 
The blue curves represent our favoured models, with $J_6$ compatible 
with \citet{Kaspi2017} but not with \citet{Kaspi2018} while the red curves are the profiles obtained from a model with a lower $J_6$, 
at the limit of what can be reached according to \citet{Kaspi2018}. Globally, the pressure and density profiles differ by a few percents at most from the ones of the
isentropic model, barely visible on the figure. However, it is worth stressing that the density of the optimized model is smaller in the Mbar region than the one of the isentropic model whereas the opposite is true in the
central regions (diluted and compact core). This is a direct consequence of the constraints arising from the gravitational moments and the Galileo observations,
as it allows to decrease the $J_4$ to $J_{10}$ values for the correct $J_2$.
In contrast,
the temperature departs from the isentropic profile for $R\lesssim 0.9\times R_J$, i.e. within
most of the interior, by a difference $\Delta T\simeq +1000$-2000 K.
Interestingly enough, this temperature increase agrees very well with the value obtained
by \citet{Fortney2003} in the H/He inhomogeneous layer for a helium enrichment in the
interior from $Y=Y_\odot$ to $Y=0.35$, and a temperature gradient leading to overstable convection. As a consequence, the 
specific entropy increases from the outer to the inner envelope. This increase is steeper
for the model with a lowered $J_6$ (consistent with  \citet{Kaspi2018}). For this latter, the inner isentropic envelope occupies
a very limited fraction of the planet, $0.89\times R_J\lesssim R\lesssim 0.92\times R_J$, and the diluted core extends up to 
85\% of the planet. Within the diluted core, the specific entropy decreases drastically due to the
strong increase in heavy elements (Figure \ref{fig:models}(b)). We do not show the 
specific entropy and compositional profiles in the diluted core because
of the degeneracy between helium and heavy element distributions in this region, which yield
similar results for the gravitational moments. Let aside the fact that the entropy profile is of no real
interest in this region. The important parameter is the steepness
of the gradient 
of composition between the inner envelope and the diluted core. The steeper the gradient, the smaller the diluted core needs to be to obtain the correct $J_2$.
The mean heavy element mass fraction is displayed in Fig. \ref{fig:models}(b). As discussed earlier, 
$Z$ is decreasing between the outer and inner envelopes ($\nabla Z>0$), as models with a continuously increasing $Z$ ($\nabla Z<0$) in the envelope, although not strictly excluded, require a very large entropy jump ($\Delta S> 1 k_\mathrm{B}/$proton), difficult
to reconcile with the examined physical processes (\S5.2.1). Future work in this direction will certainly help clarifying this issue.

\begin{figure*}[ht!]
\begin{subfigure}{.5\linewidth}
  \flushleft
  \includegraphics[width=\linewidth]{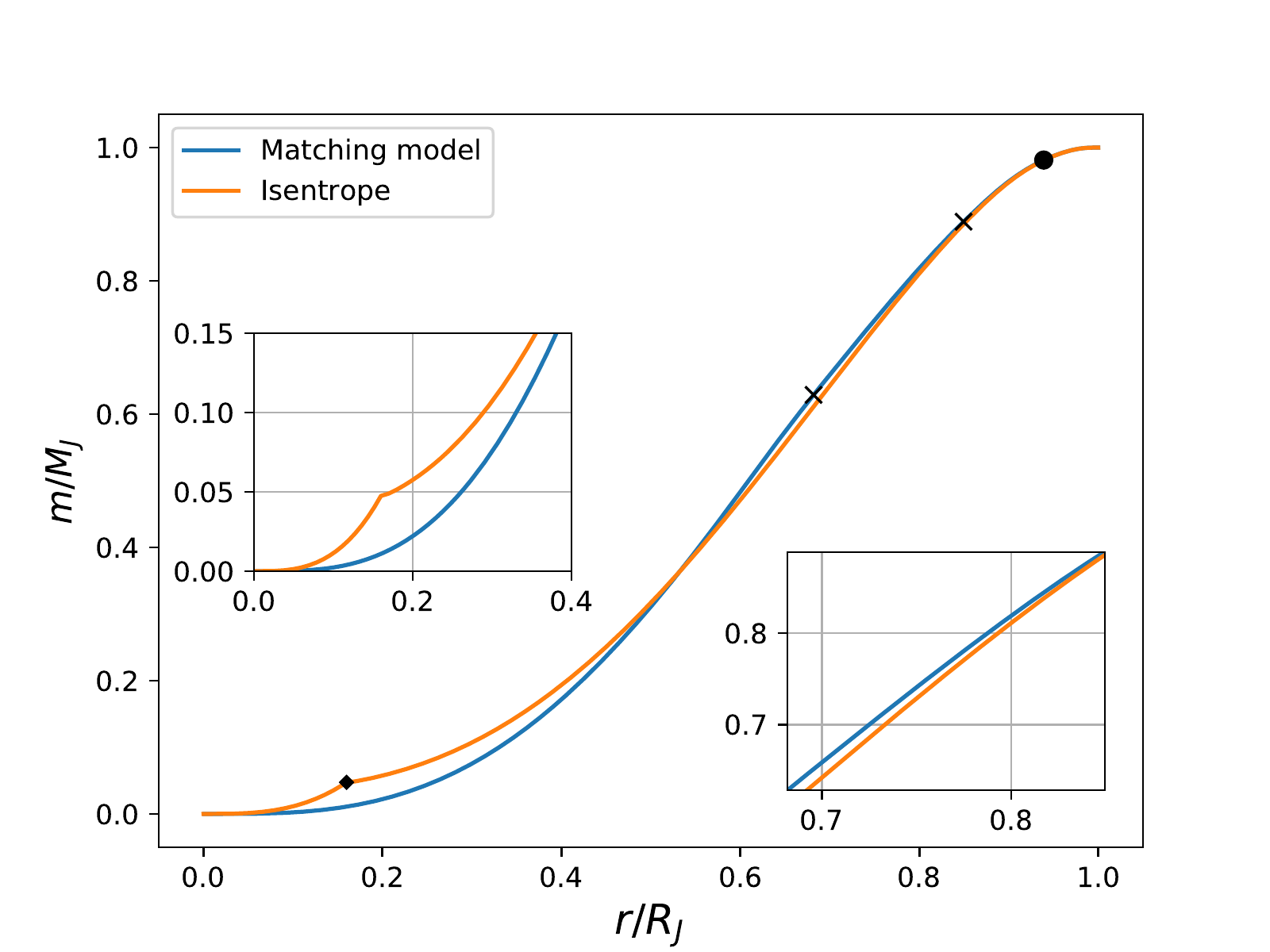}
  \caption{}
  \label{fig:mass_tot}
\end{subfigure} 
\begin{subfigure}{.5\linewidth}
  \flushright
  \includegraphics[width=\linewidth]{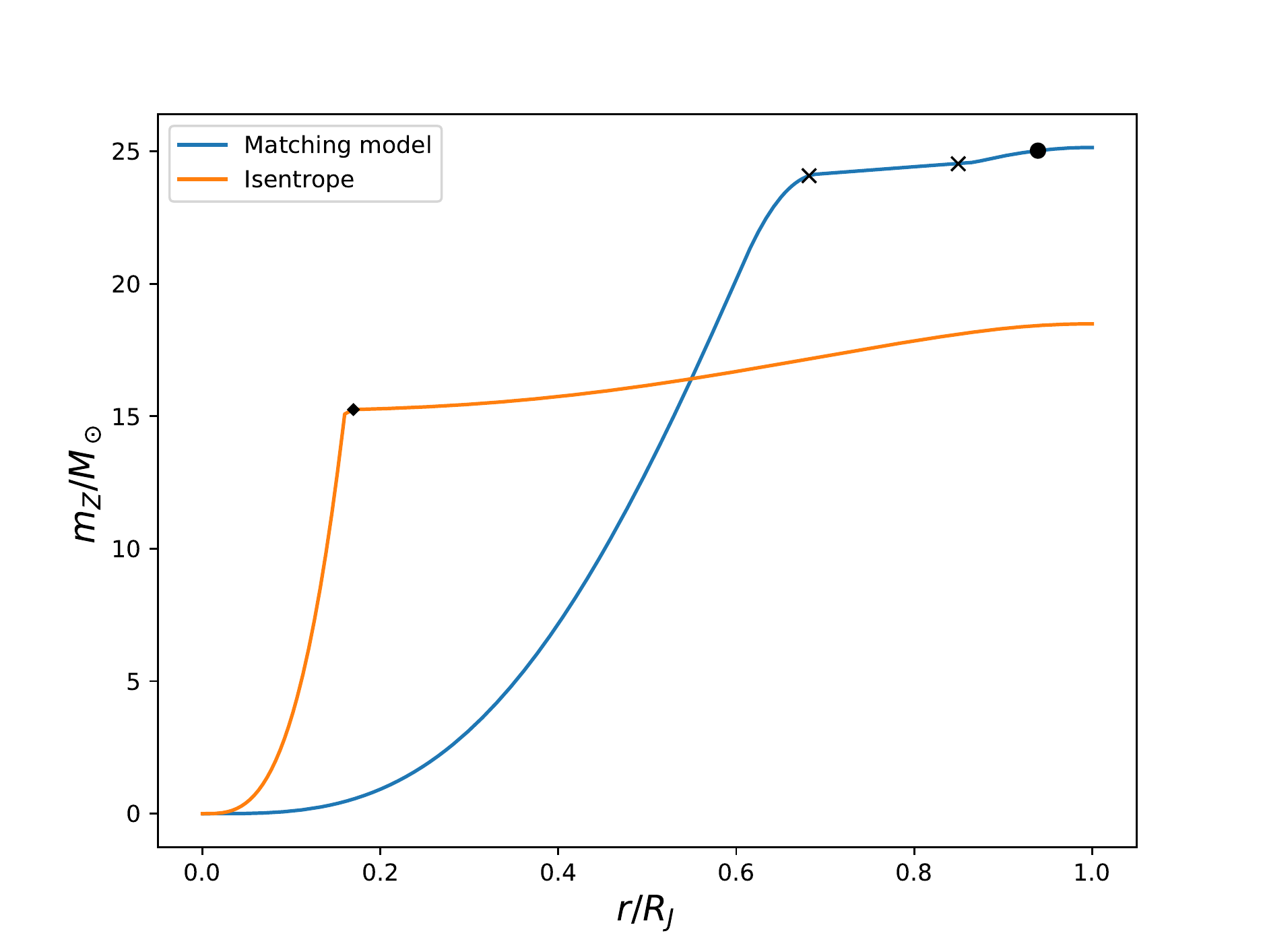}
  \caption{}
  \label{fig:mass_met}
\end{subfigure}
\caption{(a) Normalized mass as a function of radius for a model matching observational constraints
and an isentropic model. (b) Mass of heavy elements (in Earth masses) as a function of radius for the same models. }
\label{fig:mass}
\end{figure*}

Figure \ref{fig:mass}a portrays the corresponding mass profile of our typical optimized Jupiter interior structure fulfilling Juno and Galileo constraints with the wind correction of \citealt{Kaspi2017}. 
An isentropic profile is shown for comparison. 
The black circle indicates the inner limit of the outer convective zone while the two crosses bracket the inner convective zone and
the diamond corresponds to the limit of the compact core if present.
The zoom on the right hand side shows the inner convective zone, 
encompassing about $30\%$ of the mass of the planet.
 The heavy element distribution is displayed in Fig. \ref{fig:mass}b. For models with no central compact core, the total
amount of heavy elements in the planet is $M_Z=25$-30 M$_\oplus$. Adding up a compact core yields up to $M_Z=40$-45 M$_\oplus$.

\subsection{Supersolar atmospheric abundance of heavy elements}
\label{ssec:high_Z}

As discussed in \S\ref{sec:simple}, we have taken a very conservative {\it lower limit}
for the true average metal content in the atmosphere, $\zext$, in our calculations. We have taken a solar value, while Galileo 
measured abundances of individual heavy elements, excluding oxygen
and neon, rather yield $\zgal^{true}\approx (2-3) \times Z_\odot\approx 0.04$-0.06. Interestingly, such a highly oversolar value seems 
to be supported even for oxygen by the latest observations of the great red spot \citep{Bjoraker2018}. Increasing $\zext$ for a given structure
increases $|J_6|$ to $|J_{10}|$ and thus implies either a very strong 
differential rotation or a very large $\Delta S$ to preserve the moments.
If $\zext\simeq 0.05$, models with a constant inward increase of $Z$ in the metallization boundary region
lead to a $\Delta S$ much larger than the aforederived  1 $k_B/\mathrm{proton}$ maximum value consistent with
physical estimates. 
This reinforces our previous conclusion: 
\begin{itemize}
\item {\it Jupiter internal structures with an
inward increase of heavy elements within the Mbar boundary region imply uncomfortable
physical constraints:} the entropy jump or amount of differential rotation required
to be compatible with the high order gravitational moments need to be very large. In contrast, {\it models with a locally decreasing abundance of heavy element within Jupiter's metallization boundary region fulfill all constraints with acceptable levels of entropy variation and differential
rotation}. 
\end{itemize}
 
The $J$-values for five models with $\zext>Z_\odot$ are shown in Fig. \ref{fig:dS} (orange circles). We see 
that, for a given $\Delta S$, these models have higher $|J_6|$ to $|J_{10}|$ values than models with $\zext=Z_\odot$.
Although some of these models are compatible with the correction due to differential rotation estimated in \citet{Kaspi2017}, 
they are hardly compatible with the observations when considering the correction to the even gravitational moments
estimated in \citet{Kaspi2018}.
We recall, however, that all the models of Figure \ref{fig:dS} have $J_2$ and $J_4$ values consistent with \citet{Kaspi2018}. 
Because of the strong correlation between $J_4$ and $J_6$,
further decreasing $|J_4|$, consistent then with \citet{Kaspi2017} but not with \citet{Kaspi2018}, would allow us
to decrease the $|J_6|$ and higher order moment values, expanding the range of plausible models.
The derivation of precise constraints 
on the depth penetration of differential rotation and its effect on the $J$'s as a function of $\zext$ will be examined in a subsequent paper.

\section{Discussion}
\label{sec:discussion}

In this section, we examine in details the reliability of the various assumptions used 
in the models. 

\subsection{Hydrogen pressure metallization and H/He phase separation}
\label{HHe}

First, following 
the nomenclature of \citet{Stevenson1977}, 
we have assumed that Jupiter had a "hot start", 
meaning that the initial inner temperature 
of the planet was higher than the critical temperature 
of both hydrogen metallization through a PPT, $T_c(H-H_2)$, and H/He demixion, $T_c(H-He)$ (for $x_{He}=0.08$). 
According to all existing numerical simulations aimed at exploring these issues, this is quite a safe assumption. 
Further work on the metallisation of hydrogen and the H/He phase diagram will
enable us to discriminate between the sectors I, II and III of Figure 1 of these authors, namely: 

\begin{itemize}
\item Sector I : if $T_c(H-H_2)< T(P) < T_c(H-He)$, where $T(P)$ is the local temperature at pressure $P$, 
hydrogen metallisation is occuring smoothly but probably triggers H/He or Z$_i$/He immiscibility. The only possibility to
deplete (resp. enrich) the inner (resp. outer) envelope in metals is to invoke 2-body or 3-body immiscibility diagrams between partially pressure ionized heavy elements $Z^{n+}$ and neutral He, similar to what is occuring for H$^+$-He phase separation, as explored in \S \ref{thermo}, and/or external accretion events, depending on the exact value of $\zext$.
\item Sector II : $T_c(H-H_2) \sim T_c(H-He)$. As for the Sector I case, 
an inhomogeneous region
forms which is depleted in He and some Z-components, but in that case there is
a possibly of H$^+$-rich bubble nucleation and thus uplifting He-poor, Z-rich bubbles and dropping He-rich dropplets (\S\ref{sssec:1st_order_met}). 
\item Sector III : if $T_c(H-H_2) > T_c(H-He)$, 
H/He demixion has not started yet, the redistribution is due to the aforementioned H$^+$-rich, He-poor bubbles. This is probably the most unlikely situation.
\end{itemize}

These situations are imposed by the necessity to globally increase the metal content of the upper envelope (and conversely deplete the lower one)
to fulfill Galileo's constraints, $\zext\ge \zgal$, but also to enrich the inner helium content, $Y$, to balance the $Z$ decrease.
According to current work 
on metallisation and immiscibility of hydrogen and helium, even though substantial uncertainty remains, and if, as found in numerical simulations, hydrogen (or any heavy component) ionization triggers immiscibility with He atoms (or He-like ones), the Sector I case is the most likely one.
This urges the need for numerical explorations of this type of diagram and, more generally,
of the stability of H/He/Z mixtures under Jupiter internal conditions.

\begin{figure*}[ht!]
\begin{subfigure}{.5\linewidth}
  \flushleft
  \includegraphics[width=\linewidth]{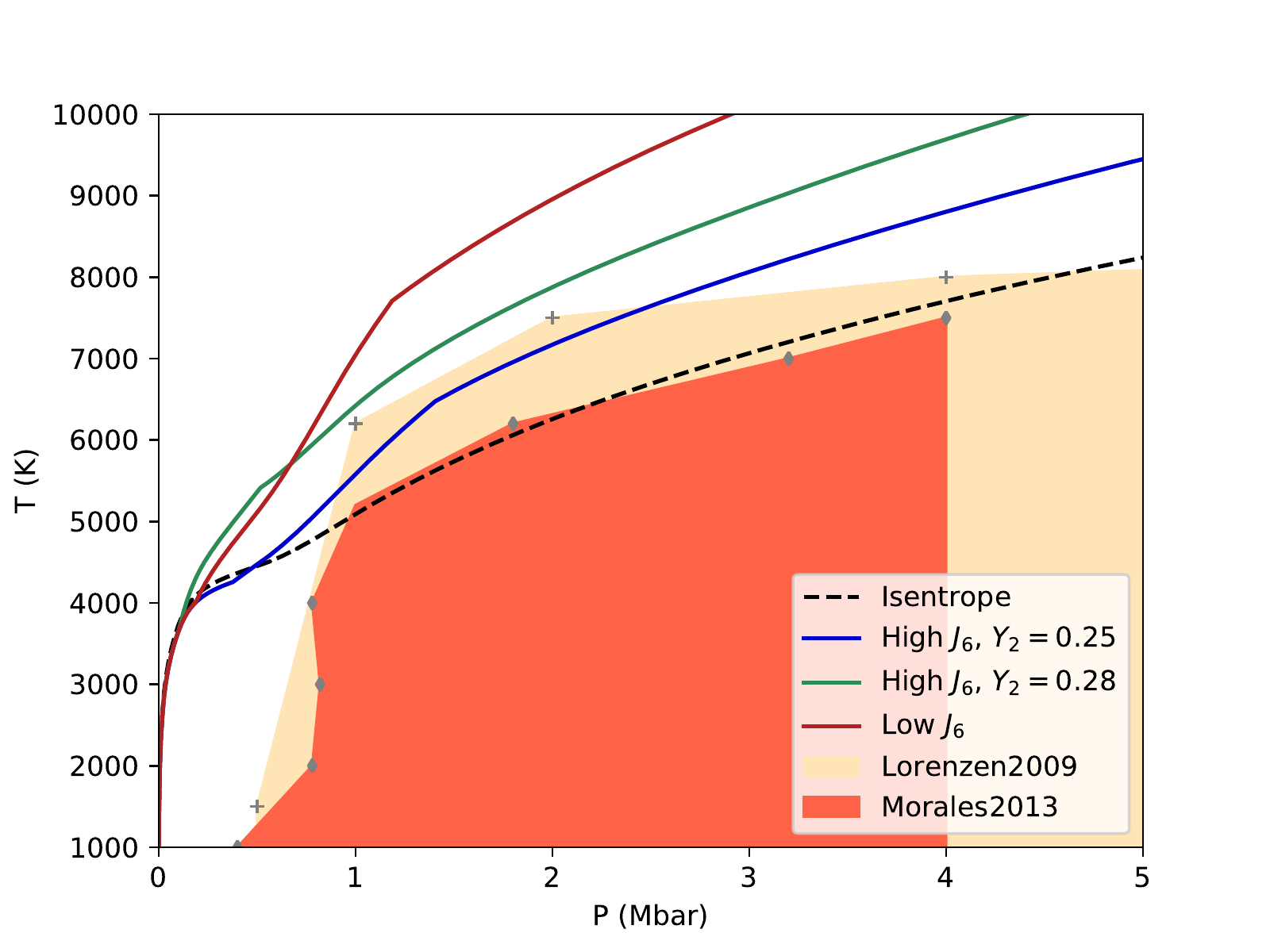}
  \caption{}
  \label{fig:normal}
\end{subfigure} 
\begin{subfigure}{.5\linewidth}
  \flushright
  \includegraphics[width=\linewidth]{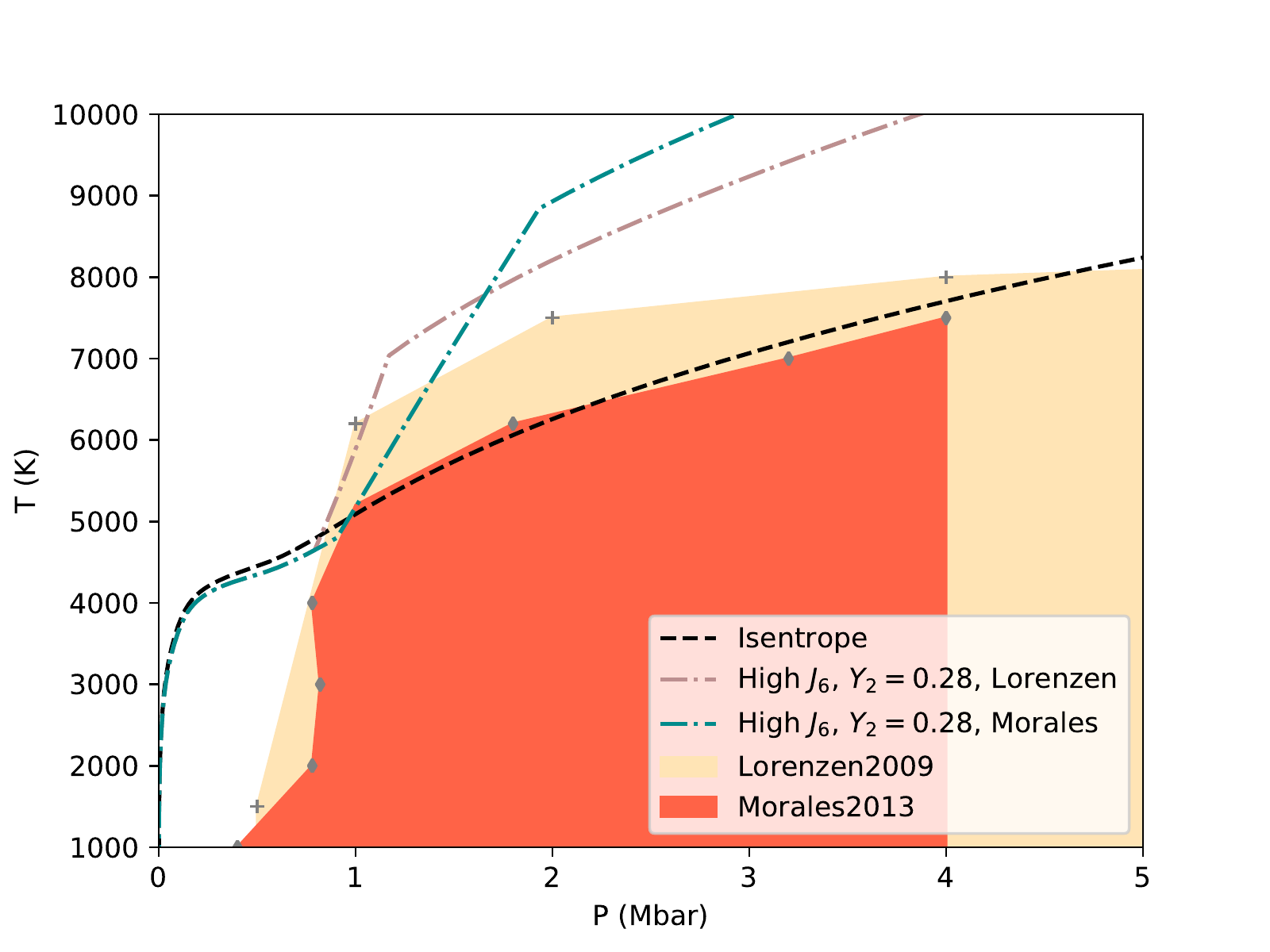}
  \caption{}
  \label{fig:low_heating}
\end{subfigure}
\caption{Temperature-pressure profiles of different Jupiter models, overplotted
on the immiscibility diagrams of \citet{Lorenzen2009} (pale yellow area) and \citet{Morales2013} ( orange area). 
All models (except isentrope) have $J_2$ and $J_4$ matching the Juno values corrected 
by the estimated differential rotation of \citet{Kaspi2018}, and a mass abundance
of heavy elements in the inner envelope $Z_2 = 0.005$. The $Y_2$ value is the mass abundance of helium in the inner envelope, 
as in Figure \ref{fig:final}. 
'Isentrope': fully isentropic structure, with $Y = 0.275$.
'High $J_6$': models with $J_6$ compatible
with \citet{Kaspi2017} but not \citet{Kaspi2018}.   'Low $J_6$': model with $J_6$ compatible
with \citet{Kaspi2018}. This latter model seems difficult to justify physically (se text).
(a) Models with layered convection fo $P\ge 0.1$ Mbar. (b) The 'Morales' and 'Lorenzen' profiles are isentropic up to the 
onset of immiscibility according to the two related phase diagrams (see text).} 
\label{fig:immiscibility}
\end{figure*}

Noticeable differences still exist between modern ab-initio calculations aimed at characterising the
H/He phase diagram. Figure \ref{fig:immiscibility} portrays the immiscibility region predicted by some of these calculations
with the $T$-$P$ profiles obtained with our favoured models fulfilling all Galileo and Juno constraints, taking
into account either the \citet{Kaspi2018} (low $J_6$) or \citet{Kaspi2017} (high $J_6$) correction due to
differential rotation. Figure \ref{fig:immiscibility}(a) corresponds to 
interior structures with strongly superadiabatic layered convection occuring at $P\ge 0.1$ Mbar. Figure \ref{fig:immiscibility}(b)
displays two models (labeled  'Morales' and 'Lorenzen', respectively) for which the change of entropy is only due
to the H/He phase separation, i.e. occurs at the corresponding critical pressures, without any 
layered convection above this layer. A Jupiter isentropic profile is shown for comparison. As seen in the figure, while, according to the
\citet{Lorenzen2009} calculations, H/He phase separation could take place in some fraction of our 
favoured Jupiter
interior models, it is not the case with the results of \citet{Morales2013} (or \citet{Schottler2018}, not shown) which predict no H/He immiscibility in present Jupiter. For the models with no layered 
convection above the phase separation (dash dotted lines in Fig. \ref{fig:immiscibility}(b)), the temperature gradient is probably too high for overstable modes to persist, 
 and convection will prevail (see e.g., Figure 3 of \citet{Stevenson1977}). 
Although the lack of excess (non-ideal) mixing entropy in \citet{Lorenzen2009} calculations casts doubt on the reliability of their phase diagram, it is worth noting
that a $\sim 200$-800 K underestimation of the critical temperature in the 1-2 Mbar domain by \citet{Morales2013} (no temperature error bar is shown in these calculations)
would be consistent with immisciblity for our model with $Y_2 = 0.25$. Therefore, Jupiter's present internal structure
could entail a region of layered convection starting around $\sim 0.1$ Mbar, associated with some change in composition, 
and a (probably small) region of H/He (most likely H/He/Z) immiscibility at deeper levels. Although more numerical exploration of this major issue is certainly needed, key diagnostics on H/He phase separation under the relevant conditions might be provided by existing experiments \citep{Soubiran2013}.

As seen in Figure \ref{fig:immiscibility}, it seems difficult to reconcile a H/He phase separation, according to the most recent calculations, with a model reproducing the \citet{Kaspi2018} $J_6$ value. Furthermore, 
the required entropy increase for this model leads to such a steep temperature gradient that unstable convection will prevail. It is thus very difficult to justify the very large entropy change $\Delta S$ in Jupiter's gaseous envelope required in this model on physical grounds. This suggests either a revision of
the \citet{Kaspi2018} analysis, or the presence of North-South symmetric winds which are inconsequential for
 the odd gravitational moments, but would increase the correction to the even gravitational moments, rejoining the corrections obtained in \citet{Kaspi2017}.

\subsection{Layered convection}
\label{layer}

As found out in the previous sections, fulfilling both Galileo and Juno constraints, while preserving a global mean helium protosolar
value $\ymean=Y_\odot$ and a physically acceptable entropy increase $\Delta S$ in the hydrogen metallization region requires an inward decrease of heavy element
abundance in this region, i.e. a locally positive gradient, $\nabla Z>0$. We verified that, because of the $\sim 1/40$ heavy element to helium number ratio, this region still exhibits a positive molecular weight gradient, $\nabla_ \mu=(d\log \mu/d\log P) >0$.
In that case, large scale adiabatic convection can be inhibited and 
lead to a regime of small scale, superadiabatic double diffusive convection (also called semi-convection) to transport heat. As mentioned previously, although a first order transition is not required to trigger such a process, it strongly favors it, as suggested for instance at the Earth's mantle boundary (e.g. \citet{Christensen1985}).

The condition for the onset of double diffusive convection reads (e.g., \citet{Stern1960}):

\begin{gather}
0 < \nabla_T - \nabla_\mathrm{ad} < \frac{\alpha_\mu}{\alpha_T} \nabla_\mu,
\label{eq:Ledoux}
\end{gather}
where $(\alpha_{\mu} = (\partial \mathrm{ln}\rho/\partial \mathrm{ln} \mu)_{P,T}$ and  $\alpha_{T} = (\partial \mathrm{ln}\rho/\partial \mathrm{ln} T)_{P,\mu}$.
In geophysics, it is well known that double-diffusive convection generally takes the form of oscillatory convection or layered convection, i.e. a stack of small-scale convective layers of size $l$ separated by diffusive interfaces (e.g., \citet{Rosenblum2011}).
In astrophysical objects, because essentially of the lower Prandtl number, this layering is more blurred
and, according to simulations, double-diffusive convection rather takes the form of ill-defined turbulent interfaces, even though finite amplitude layering remains a possibility \citep{Moll2016}.
In the absence of simulations in the present context, we will use the analytical formalism derived by \citet{Leconte2012} to verify the presence 
of layered convection in our models. As shown by these authors, this is controlled by 
the parameter $\alpha$, which is the ratio of the size of the convective layer
to the pressure scale height, $\alpha=l/H_P$. From their eqn.(21), we can relate
this parameter to the superadiabatic gradient, $(<\nabla_\mathrm{T}> - \nabla_\mathrm{ad})$, by: 

\begin{equation}
<\nabla_\mathrm{T}> - \nabla_\mathrm{ad}= \epsilon_d \times \Bigl[(\Phi_0 \alpha^4 \epsilon_d)^{\frac{-1}{4(1+a)}} + 
(\Phi_0 \alpha^4 \epsilon_d)^{\frac{-a}{(1+a)}} \Bigr],
\end{equation}
with layered convection occuring when 

\begin{equation}
10^{-9}-10^{-6}  \lesssim \alpha \lesssim 10^{-4}-10^{-2},
\label{eq:alpha}
\end{equation}
with, for Jupiter, the lower bound being more likely $10^{-5}$ (see Table 1 of  \citet{Leconte2012}).

\begin{figure*}[ht!]
\begin{subfigure}{.5\linewidth}
  \flushleft
  \includegraphics[width=\linewidth]{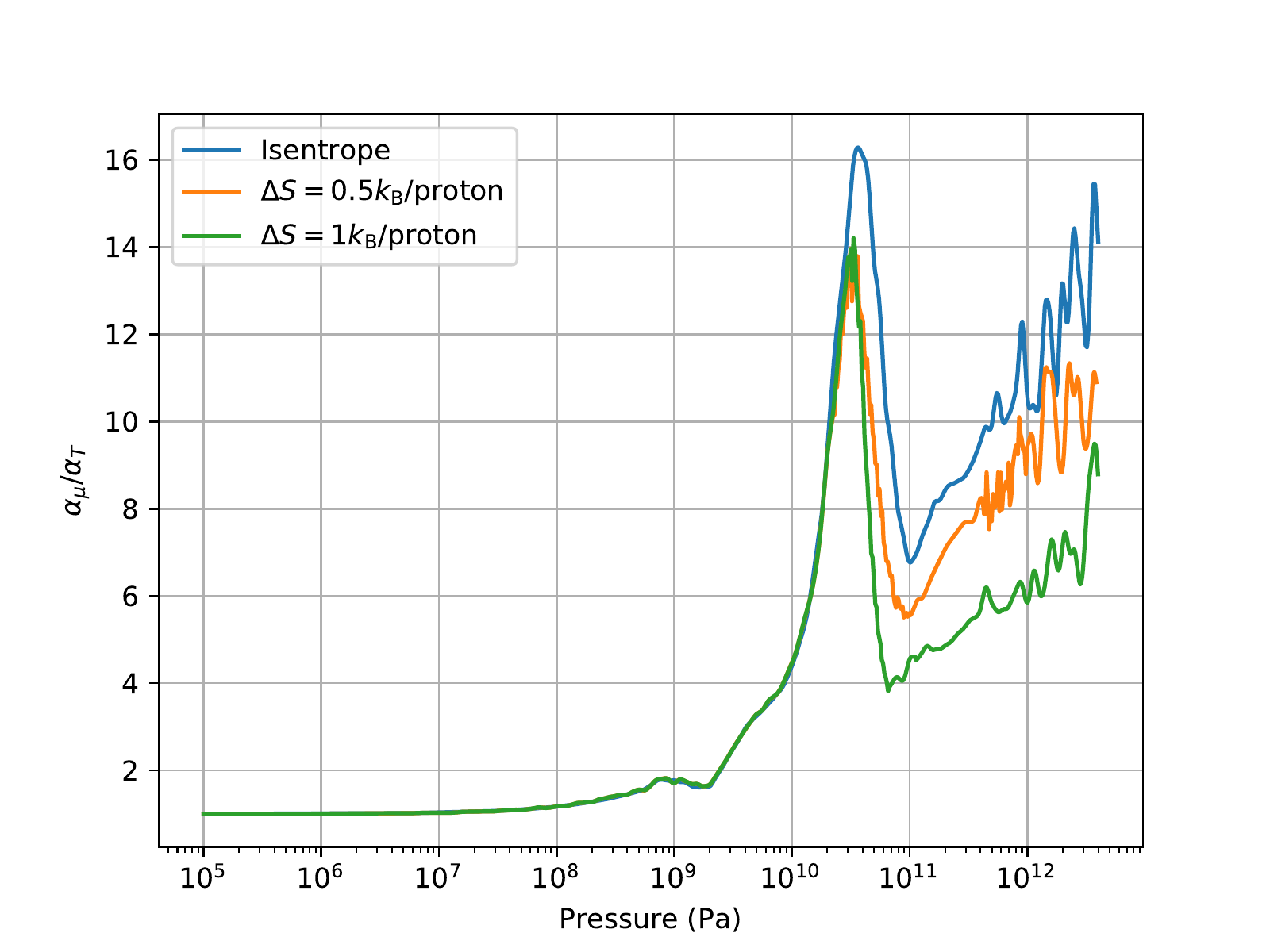}
  \caption{}
  \label{fig:alpha_LC}
\end{subfigure} 
\begin{subfigure}{.5\linewidth}
  \flushright
  \includegraphics[width=\linewidth]{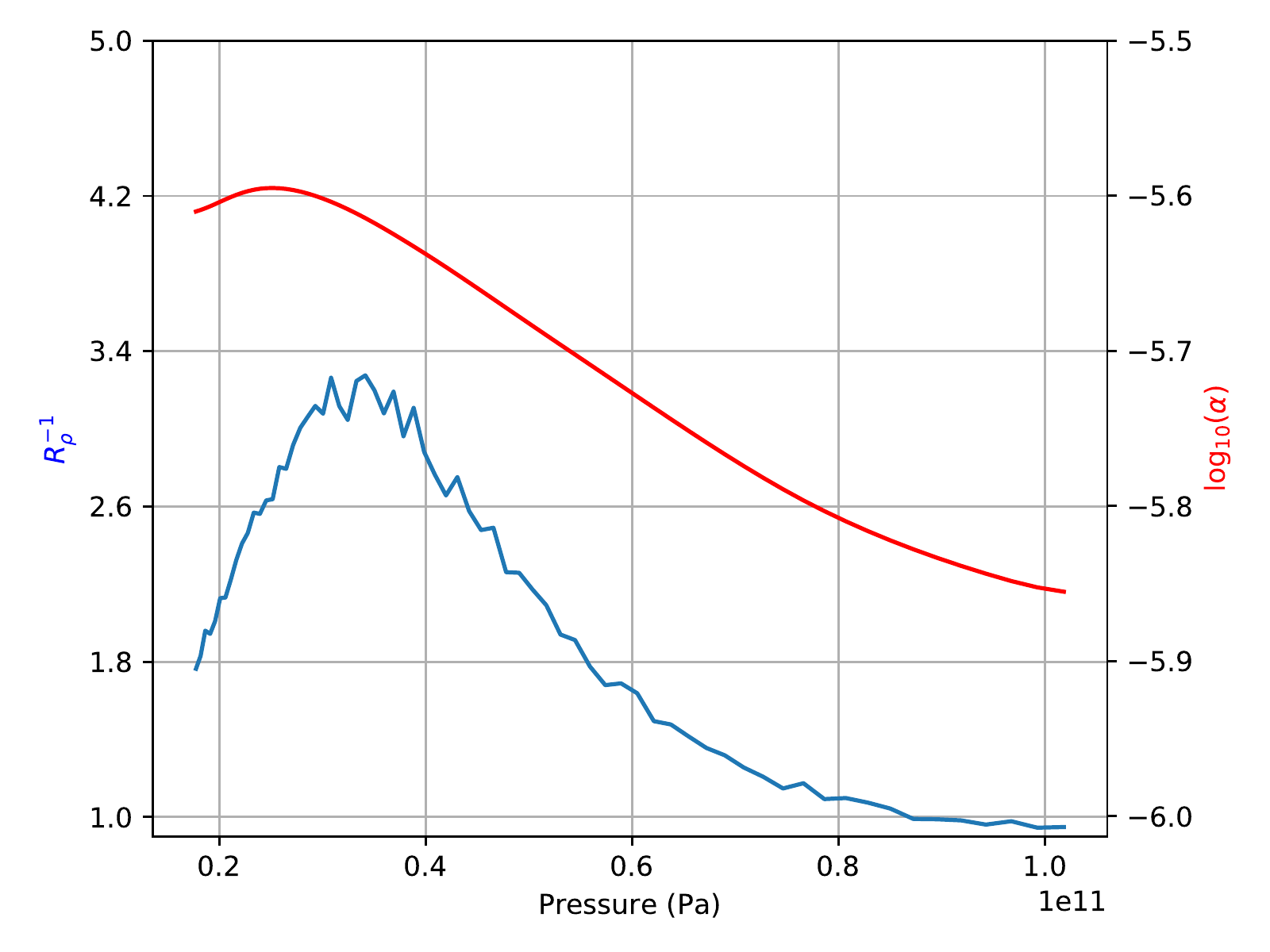}
  \caption{}
  \label{fig:Ledoux}
\end{subfigure}
\caption{(a) $\alpha_{\mu}/\alpha_T$ as a function of pressure for an isentropic model and 
two models with $\Delta S =$ $0.5$ and $1\,\mathrm{k_\mathrm{B}/proton}$, respectively. (b) $\alpha$ (eq.\ref{eq:alpha}) and $R_\rho^{-1}$ (see \citet{Leconte2012}) for the model described in the text.  }
\label{fig:LC}
\end{figure*}

As MacLaurin spheroids have by definition a constant density, layered convection cannot be prescribed very accurately with the CMS method.  
As for the case of a first order 
phase transition/separation, we have implemented a sharp entropy and composition change at constant $T$ and $P$
between consecutive layers.
We can then verify, for the appropriate models, whether
conditions (\ref{eq:alpha}) and (\ref{eq:Ledoux}) are fulfilled or not.  
Figure \ref{fig:LC}a displays the values of $(\alpha_{\mu}/\alpha_T)$,  calculated with our EOS, for an isentropic 
profile and for two profiles with an entropy increase 
in the Mbar region of $0.5$ and $1\,\mathrm{k_\mathrm{B}/proton}$, respectively. We see that this quantity increases with depth by an order of magnitude, between
$\alpha_{\mu}/\alpha_T=1$ in the external layers and the values prevailing at depth in Jupiter, due essentially to the onset of of H$_2$ dissociation (see \citet{Chabrier2018}).
This favors the onset of layered convection deeper than $\sim0.1$ Mbar (see eqn.(\ref{eq:Ledoux})). Figure \ref{fig:LC}b 
displays the values of $\alpha$ and of the parameter $R_\rho^{-1} = (\alpha_{\mu}/\alpha_T) \nabla_\mu/(\nabla_T - \nabla_\mathrm{ad})$ (overstable convection occurs for $R_\rho^{-1}>1$, see \citet{Rosenblum2011}, \citet{Mirouh2012}, \citet{Leconte2012}),
for a composition change from ($Y_{ext}=0.23  \ , \zext=0.03$) to ($Y_2=0.3  \ , \ Z_2=0.01$), with an 
entropy increase $\Delta S=0.45\, \mathrm{k_\mathrm{B}/proton}$ between $0.1$ and $1$ Mbar. 
We see that $\alpha$ and $R_\rho^{-1}$ fulfill the conditions for the presence of layered convection in this domain. Models with higher $\Delta S$ (of at most $\sim 0.6\, \mathrm{k_\mathrm{B}/proton}$, see \S\ref{ssec:entropy_physics}) require larger $\Delta Y$. Globally, we verified
that all our favoured models do fulfill the
conditions for the occurence of layered convection derived in  \citet{Leconte2012}. 

One word of caution should be noted: when H$_2$ dissociates into atomic H$^+$, 
the mean molecular weight $\mu$ decreases brutally.
According e.g. to \citet{Nellis1995}, however, the fraction of dissociation is about $10\%$
at $1.4$ Mbar. The molecular weight thus remains barely affected up to this pressure and the decrease of $\mu$ do to H$_2$ dissociation should happen over a rather limited region
between $\sim1.4$ and $2$ Mbar. Whether layered convection is still present or not in this domain is less clear (although overshoot probably occurs) but we consider
it to be localized enough to not significantly modify the aforecalculated temperature profile.

\subsection{External impacts, atmospheric dynamical effects}
\label{dynamic}

We examine  here whether Galileo's high heavy element abundances in the outermost part of the envelope could be due to recent impact or recently accreted material which would not
have had time yet to be redistributed within the planet and thus would not affect its gravitational potential. In that case, Juno's contraints
could be examined without taking into account the ones from Galileo. Using standard equations of the mixing length 
theory (\citet{Kippenhahn1990}, \citet{Hansen1994}), the typical convective velocity in Jupiter scales
as:
\begin{equation}
v_\mathrm{conv} \approx 10\,({\nabla-\nabla_{ad}\over 10^{-8}})^{1/2} ({H_P\over 10^6 \,{\rm m}})\,\,{\rm m}\,{\rm s}^{-1},
\label{eq:conv}
\end{equation}
with $(\nabla-\nabla_{ad})$ ranging from $\sim 10^{-10}$ at the center to $\sim 10^{-6}$ near the surface, yielding $v_\mathrm{conv}\sim$ a few to about $\sim 100$ cm s$^{-1}$ from the center to the surface.

This means that, within at most a few
years, the external extra material will be mixed throughout most of the planet. It is quite clear that Jupiter has 
not accreted a few Earth masses of heavy elements in the past 20 years. The only source of uncertainty is Shoemaker-Levy 9. Although 
its mass is ridiculously small compared to the mass of the envelope, and even to the mass of heavy elements in the envelope of Jupiter, the energy deposited
when it crashed onto the planet (end of July 1994) triggered uplifting of deep material (e.g, \citealt{Bezard2002} or \citealt{Moreno2003}). As Galileo entered Jupiter's atmosphere 
1.5 year later (7 December, 1995), the material had
probably be mixed again throughout the upper envelope (remember  $ v_\mathrm{conv}\approx$ 1 $\mathrm{m\,s}^{-1}$ in this region).
Therefore, taking into account Galileo's observations of
Jupiter's heavy element external abundance seems to be mandatory when trying to recover Juno's gravitational moments.

Atmospheric dynamical effects (for instance a vortex), on the other hand, could have produced a localised maximum concentration of heavy material, not representative of the (lower) average value. 
Additional information is provided by the Juno microwave instrument \citep{Li2017}
that suggests a rather unexpected amonia
vertical profile. However, if we compare the Fig. 4 of these authors
with the location where the Galileo entry probe dived (around 5-10$^\circ$ in planetocentric
latitude), we see that, compared to the region deeper than $100\, \mathrm{bar}$,
Galileo can have only measured a lower limit of the amonia content. 
Indeed, there seems to be a maximum of amonia concentration in the equatorial regions, 
followed by a minimum in the mid latitudes. Galileo entry probe dived at the limit between
these two regions, and, most importantly, the concentration anywhere above the $25$ bar level
 is smaller than the one deeper
than $100$ bars, where convective mixing definitely takes place. Therefore it seems unlikely 
that the measurements of Galileo, consistent with an increase of ammonia with depth up to 
$25$ bars, are an upper bound of the heavy element composition.

One can also wonder whether the composition varies between the $P=1$ bar and
$P = 100\, \mathrm{bar}$ levels, affecting the gravitational moments, 
and then should be parametrized instead of being assumed to be constant.
Between two models with $(Y,Z) = (0.23,0.02)$ and $(Y,Z) = (0.23,0.04)$ 
in the external envelope, we get changes in $J_2$ and $J_4$, due to the first 100 bar variation,
of at most $2 \times 10^{-6}$ and $2 \times 10^{-7}$, respectively, about an order of magnitude smaller than the change 
due to differential rotation (Fig. 3 of \citet{Kaspi2017}). Variations of composition in the external envelope are thus a 
second order correction compared to differential rotation. 

\subsection{Magnetic field}

Although the magnetic field at the surface of Jupiter has been shown to vary with 
latitude and longitude within an order
of magnitude \citep{Connerney2018}, the leading feature is a dipolar field with moment $M = 4.170\times 10^{-4}\, \mathrm{T}$.
According to numerical simulations, self consistent dynamo action is generally found to start when the convective magnetic Reynolds number,
i.e. the ratio of magnetic field production to Ohmic
dissipation, $Rm=v_{rms}d/\eta$, where $v_{rms},d$ and $\eta$ denote respectively the rms flow velocity, the thickness of the shell and the magnetic
diffusivity, exceeds a critical value $Rm\gtrsim 50$ \citep{Christensen2006}. Using simple scaling arguments (e.g. \citet{Chabrier2007_2} and
\S\ref{dynamic}),
it is easily verified that this condition is well fulfilled at the ionization boundary, located around $P \approx 1 \,\mathrm{Mbar}$, 
where the density is $\rho \approx 800\, \mathrm{kg\,m}^{-3}$, 
radius $R \approx 0.85\times R_\mathrm{J}$, and that deeper in Jupiter convective, metallic zone we get $Rm\gg 10^5$. This suggests that 
the primary dipole-dominated magnetic field is created at depth where $Rm$
is significant, the electrical conductivity high, and the density contrast relatively mild. This is indeed what is found in state-of-the-art
numerical simulations which reveal that the combination of a deep-seated dipolar dynamo and a magnetic banding associated with the
equatorial jet reproduce Jupiter field geometry with realistic relative axial dipole, equatorial dipole, 
quadrupole and octupole  field contributions \citep{Gastine2014}.
These simulations are also consistent with the suggestion that 
the mean internal field strength as well as the mean convective velocity
scale with the available convective power \citep{Christensen2006}.
\citet{Gastine2014} find that Jupiter's surface magnetic field strength, $B_J\simeq 7$ G, is consistent with a typical rms flow 
velocity $\sim 3$ cm s$^{-1}$, for a shell thickness extending from 0.2 to 0.99 $R_J$. Such a velocity is
largely consistent with the maximum value derived in \S\ref{dynamic} around the metallization boundary. Although a dedicated study
is necessary to explore this issue in the presence of an outer layered convection region, the rms velocity and the average conductivity
should remain large enough for $Rm$ to still exceed the critical value $R_m \approx 50$, and thus for the reservoir of convective power to still contribute
appreciably to the dynamo action.

Defining $R_{50}$ as the radius in the planet above which $Rm\lesssim 50$, \citet{Duarte2018} find that
$R_{50}\simeq 0.9\,R_J$, due essentially to the big change of conductivity when molecular H$_2$ fully recombines, while values 
 below $R_{50}\simeq 0.85\,R_J$ seem to be excluded with some confidence. This is consistent with our favorite models (Fig. 6).
 without inclusion of the region of compositional change. 
 
Interestingly enough, recent observations of the hemispheric difference in Jupiter's field geometry \citep{Moore2018} are consistent with the superposition of
two types of dynamo action, namely a thick shell dynamo, reponsible for the strong axial dipole, occuring at depth in the metallic region, and
a thin-shell dynamo, yielding the observed hemispheric asymmetry, occuring further up in the envelope. A region of density gradient/layering between
these two regions provides a very plausible explanation for such a separation.

Note that the large-scale field generation also constraints the size and or the magnetic/electric properties of the diluted core. In case this latter is unable to sustain dynamo action, this implies that the inner convective envelope 
must be sufficiently large to generate the magnetic field, limiting the maximum extension of
the diluted core. This in turn
 limits the maximum mass of the central dense core. Indeed, as shown in the previous sections, the bigger the compact core the smaller
 (in absolute value) the high order gravitational moments but the bigger the diluted core.
 If, however, the conductivity in the diluted core is large enough to generate electric currents, flow motions generated by 
 density contrasts (due e.g. to
 ill-mixed elements) and the Coriolis force resulting from Jupiter's spin might be able to produce magnetic fields and sustain a geodynamo process similar to
 the one taking place near the Earth iron core.
In that case, the
 diluted core might contribute substantially to the field generation, extending the available domain for global dynamo action.
We realyze that at this stage such a discussion is purely speculative but we hope to motivate dedicated explorations of these issues as
Jupiter's magnetic field generation can certainly help assessing the reliability of the present structure models.

\subsection{Evolution}
\label{evol2}

Constraints due to Jupiter's evolution have been briefly examined in \S\ref{evol}. Our favoured models suggest an entropy jump
$\Delta S\approx 0.3$-1 k$_B$/proton between the outer molecular envelope and the inner metallic one (see Fig. \ref{fig:dS}), which yields a warmer inner temperature for the
planet than in the absence of $\Delta S$.
This temperature difference, due to the entropy gradient in the inhomogeneous region, will keep increasing with time as either layered convection and/or H/He phase separation and helium sedimentation will keep progressing. This yields a heat release from inside to outside during the planet's evolution.
Jupiter's observed luminosity today, however, suggests that, if H/He phase separation does occur in the planet, it must contribute only modestly to its cooling history. This condition can be fulfilled for several reasons. (1) If a significant fraction of this energy is devoted to
heating up the interior (keeping the inner convective envelope on a high isentrope), the energy release, whatever its source, remains modest along cooling. Not mentioning the fact that H/He might become miscible again. (2) H/He separation in Jupiter may have started only recently, contributing negligibly to Jupiter's luminosity (power) whatever the
He sedimentation energy release. (3) The H/He immiscible region, yielding a
temperature gradient, might encompass a relatively modest fraction of Jupiter's interior.
(4) More interestingly in the context of our favoured models, core erosion, if it occurs, implies that a fraction
of the planet's internal energy must be consumed in the redistribution of heavy elements against gravity, and thus be transformed in
potential energy (see e.g. \citet{Stevenson1985}, \citet{Guillot2004}). This consumption of Jupiter's available internal
energy will speed up the cooling of the planet. Even in the presence of layered convection, the final energetic balance might eventually decrease or
increase the planet's cooling rate \citep{Leconte2013}. 

In summary, if the present final models are representative of Jupiter's present internal structure and composition, its cooling history
should include (i) layered convection, (ii) H/He (or other elements) phase separation plus helium dropplet sedimentation and (iii) core erosion. Finding
out what will be the resulting impact of these three processess upon the planet's global cooling history is a highly non trivial task, which
can hardly be intuited or inferred with simplistic models.

\subsection{Does the observed outer condition lie on an adiabat ?}

In this section, we raise an other issue regarding Galileo's constraints.
Galileo's measurements are taken from 1 to about 25 bar and in all existing models, including the ones derived in the previous sections,
the temperature profile is supposed to follow an isentrope, starting from the observed value $165$ K at 1 bar (for reversible processes like convection, an adiabat is equivalent to an isentrope, $dQ=TdS=0$).
It is not obvious, however, that the deeper profile (between say 1 and 100 bars) does follow an isentrope. The measurements of Galileo
show an increase of heavy element abundance with depth, indicating that, at $P\simeq 25$ bar, the probe
has not reached yet a well mixed region. In case of departure from adiabaticity,
the outermost temperature gradient could then be larger
than the isentropic one, implying that the real inner entropy profile lies on a warmer isentrope than the one obtained if one assumes
it is given by the $P=$1 bar, $T=165$ K condition.

Let us consider, notably, the impact of rotation.
There is presently no well defined theory for turbulent convection in the presence of rotation so we can only rely on estimates. At the pressure level $P\simeq 10$ bar in Jupiter external envelope, the optical depth is $\tau\gg 100$, so except for the possible impact of rotation,
one can safely assume that the profile is isentropic at this level.
This pressure corresponds to $\rho \sim 1$ kg m$^{-3}$, $g=GM_J/R^2\simeq 20$ m s$^{-2}$,
$T\sim 200$ K, and thus a typical convective length $l\sim H_P=P/(\rho g)\approx 10^5$ m, about 1\% of
the planet's radius.
Assuming all Jupiter internal flux, ${\mathcal F}\simeq 5.4$ W m$^{-2}$, is transported by convection and using the usual equations of the Mixing Length Theory \citep{Kippenhahn1990}, this yields a typical superadiabatic gradient $\Delta\nabla T_0=(\nabla-\nabla_{ad})_0\approx 10^{-6}$
at this pressure level, i.e. a characteristic convective velocity $v_{conv} \lesssim 1$ m s$^{-1}$.
Since Jupiter angular velocity is $\Omega_J=v_{rot}/R_J=1.76\times 10^{-4}$ rad s$^{-1}$, the ratio of inertial to Coriolis
forces, known as the Rossby number, at 10 bars is thus $Ro=v_{conv}/(l\Omega)\gtrsim 0.1$. Convection at the top of the upper envelope, where the
Galileo measurements have been made, should thus be only moderately affected by rotation. 
It is also easily verified that the Coriolis acceleration is much smaller than the gravity, $R_J\Omega^2\ll g$, which allows to perform a linear stability analysis of the MLT equations in the presence of a Coriolis force, $2\Omega {\bf \times} v_{conv}$, \citep{Chandra1961}. Additionally, given the value of the Rossby number, this linear analysis can be performed in the strong rotation limit \citep{Stevenson1979}.
This yields for the suradiabtic gradient in the presence of rotation in the region probed by Galileo \citep{Stevenson1979}:

\begin{eqnarray}
\Delta\nabla T_\simeq (\Delta\nabla T_0)^{3/5}\left(\frac{\Omega^2 l}{g}\right)^{2/5}\simeq 6\times 10^{-6}.
\end{eqnarray}

This estimate shows that rotation cannot yield a strong departure from adiabaticity in the outermost envelope layers of Jupiter, as
expected from the inferred Rossby number value, in
contrast to deep convective regions (see e.g. \citet{Chabrier2007_2}).

Moreover, at a pressure of 1 bar, the atmosphere is composed of alternative superrotating and underrotating stripes in latitude. Both at the equator, where advection dominates, and in the mid to high latitudes, where geostrophy applies, one can show that the horizontal variation of temperature due to the winds is of the order of a few percents in latitude and longitude, with a maximum at the equator (as confirmed by the observations of \citet{Fisher2016} and the temperature profiles from GCM calculations of \citet{Schneider2009}). In that regard, the measurement of Galileo are rather an upper bound than a lower bound on the temperature, and deviations from these measurements are small. 

In conclusion, it seems quite safe to assume that the external condition defined by the $T=165$ K,
$P=1$ bar condition lies on an adiabatic profile.

\section{Conclusion}
\label{sec:conclusion}

In this paper, we have examined models of Jupiter aimed at fulfilling {\it both} the most recent
Juno observations and the Galileo constraints. Our calculations were carried out with the Concentric MacLaurin Spheroid method with all
the necessary mathematical and numerical constraints \citep{Debras2018}. Because of the 
tension due to the high observed abundances of helium and heavy elements 
in the external envelope and the low values of the high-order gravitational moments, 
the number of possible interior models is very limited. We
verified that, even if 
the 1 bar temperature observed by Galileo departs from an adiabat because of the impact of rotation, the departure remains modest enough
to take the $P=1$ bar, $T=165$ K observed values as the external isentrope conditions.

We first showed that the new data from Juno cannot
be reproduced with conventional 2- or 3- isentropic homogeneous layer models. These latter are not 
able to match both the values of the gravitational moments and the 
external abundance of metals, which confirms 
the analysis of \citet{Wahl17}. The first conclusion is that there must be at least two regions of compositional gradient within the planet's interior. 

Our thorough analysis suggests that the planet should be composed of at least four main regions, namely, moving inward from the surface: (I) the external isentropic, homogeneous molecular/atomic H$_2$/He/Z envelope, extending downward to about $93 \%$ of the planet's radius, 
(II) an inhomogeneous domain around $P\sim 0.1$-2 Mbar, encompassing the region of hydrogen pressure ionization, of size about $\sim 10-15\%$ of the radius,  characterised by a gradient of composition ($\nabla X, \nabla Y, \nabla Z$), and an inward positive entropy change, $\Delta S>0$ (i.e. a locally negative entropy gradient, $\nabla S=(\partial S/\partial r)< 0$),
(III) a second inner isentropic, homogeneous, metallic envelope hydrogen, extending from the bottom of region II down to $60-70 \%$
of the radius, lying on a hotter isentrope than the outer envelope one ($S_{III}>S_I$) with, most likely, a smaller metal mass fraction than
in the outer homogeneous enveloped ($Z_{III}<Z_I$), and (IV) a diluted $Z$-rich core composed of volatiles, exhibiting a compositional gradient. Potentially, a central compact seed can be present, essentially composed of solid iron and silicates. 

A major result of this study is that a substantial entropy increase, $\Delta S\gtrsim 0.3\, \mathrm{k_B/proton}$ is necessary in the
inhomogeneous region II to fulfill all the observational constraints. If not, one needs to invoke very strong differential rotation to explain the values of the high order gravitational moments, at odd with the estimate of \citet{Kaspi2018} (and 
even \citet{Kaspi2017} if $\Delta S = 0$). 
This suggests the occurence of either superadiabatic layered convection and/or a first order phase transition, be it hydrogen pressure ionization or H/He phase separation.
If this entropy increase lies in the range $0.3\lesssim \Delta S\lesssim 1\, \mathrm{k_B/proton}$,
which seems to be inferred from various relevant physical processes, the abundance $Z$ of heavy elements in region II must be locally {\it decreasing}, i.e. exhibiting a {\it positive gradient of composition}, $\nabla Z>0$, but an increasing molecular weight i.e. a negative molecular weight gradient, $\nabla \mu <0$, due to the much larger helium fraction at the bottom of region II. 
In case of a strong entropy increase in region II, $\Delta S>0$, 
it is possible to 
 strongly reduce the value of the high order gravitational moments while still fulfilling Galileo's external metal abundance,
by invoking the presence of a central compact core. The first impact of this latter is to restrain the mass domain of Jupiter's interior
impacting the moments. Although it is possible to find models with an inward increasing metal abundance within region II, $\nabla Z<0$,
 compatible with Juno and Galileo, they require such a large $\Delta S$ or amount of differential rotation that it seems hardly possible to justify them on physical grounds.

Note that there is a degeneracy of solutions between the change of
entropy $\Delta S$ in region II and the outer differential rotation. The stronger $\Delta S$
the shallower and weaker the differential rotation, enabling eventually values consistent with the estimate of \citet{Kaspi2018}. In contrast, 
if differential rotation extends deeper inward and/or is stronger than suggested by these authors, the change 
of entropy across the boundary region can be significantly lowered. According to the study of \citet{Cao2017}, however, the differential rotation cannot extend too deep, as magnetic reconnection eventually occurs deep in the envelope, leading to rigid rotation.

The entropy jump $\Delta S$ in region II is also related to, and can be constrained by other conditions.
Namely, (i) the mass of the central dense core $M_c$. Indeed, as shown in the study, 
the mass of the central core is directly correlated with $\Delta S$ (the larger $\Delta S$ the larger
$M_c$)
and then anti-correlated with the amplitude of the high order 
gravitational moments. 
(ii) The gradient of helium and heavy elements within the boundary region II: the larger 
the increase in Y and Z between region I and III (most probably an increase in Y and a decrease in Z), the larger the $\Delta S$ required to reproduce Juno's data. (iii) At last, 
$\Delta S$ is constrained by the
physics of dense matter, namely the nature of hydrogen pressure ionization (critical temperature and pressure and entropy discontinuity) and
by the miscibility diagram not only of H/He but also of the various dominant heavy elements in metallic hydrogen.
Finally, it is worth pointing out that, even in the absence of a first-order transition, region II, characterised by a strong
compositional change, is prone to layered convection. As examined in \S \ref{sec:discussion}, the inferred profile is indeed consistent
with conditions derived in \citet{Leconte2012} for the presence of layered convection.

The inward decrease of the mean heavy element mass fraction in region II, and thus the oversolar value in the upper envelope inferred from 
Galileo, can have
different explanations. In case the local temperature at the H$_2$-H$^+$ metallization pressure $P_c$ is lower than
the critical (PPT) temperature ($T<T_c$), nucleation of H$^+$-rich bubbles can occur, associated with some heavy elements, and move upward
across the critical line, enriching continuously the upper envelope I in (some) heavy elements. Since hydrogen ionization immediately triggers
H/He phase separation, with the formation of drowning He-rich dropplets, this process yields at the same time an enrichment of He and
associated species in the lower envelope III. In case the above temperature condition is not fulfilled, hydrogen pressure ionization occurs
smoothly, there is no bubble nucleation. In that case, in order to enrich the upper envelope I in heavy elements there must be either an immiscibility of some species in the H/He/Z mixture at the relevant temperature and pressure, yielding a large equilibrium concentration of these species in the low pressure, low temperature molecular phase, or persistent layered convection.
If the real enrichment in heavy elements is largely oversolar (i.e. much larger than Galileo's value) the occurence of
external impacts during Jupiter's history seems to be inevitable to explain it.
In all cases, it seems difficult to avoid the presence of a first-order transition or persistent superadiabatic layered convection in Jupiter's
gaseous envelope around the $\sim$ Mbar region. Accurate calculations of the planet's long term evolution are definitely needed to
assess or reject the viability of some of these static models. As mentioned in \S\ref{evol2}, however, properly handling such calculations appears to be a task of enormous complexity.

In conclusion, we have derived in this study interior models of Jupiter able to
reproduce {\it all} the observed stringent gravitational constraints from the Juno mission 
and the outer helium and heavy element abundances from Galileo. 
These models differ appreciably from all Jupiter model derived previously, which ignored either Juno or Galileo's constraints, making 
these models (and related papers or reviews) obsolete. 
As shown above, however, because of the lack of precise characterisation of major physical
processes, there is still a degeneracy 
of possible models. Indeed, neither experimental nor numerical explorations of these processes provide yet definitive information about the related fundamental
questions. This illustrates the tight link between fundamental physics and astrophysics.
Additional constraints also arise from the differential rotation in the planet. Indeed, high order gravitational moments are essentially
 only sensitive to 
the outermost region of the planet, constraining the available level of differential rotation (see notably \citet{Hubbard1999}). As explained above, more constraints on differential rotation  will help
constraining the change of entropy in the pressure ionization boundary domain, and subsequently the mass or even the presence of the central compact core. This issue will be explored in a forthcoming paper. At last, it is worth mentioning that the most favoured models
able to fulfill both Galileo's and Juno's constraints, according to the present study, are basically the ones, or at least among the ones
 intuited and explored in great details by Stevenson \& Salpeter 40 years ago in their two seminal papers and by \citet{Stevenson1985} !

\subsection*{Acknowledgments}

We wish to thank T. Fouchet, E. Jaupart, G. Laibe and D. Stevenson for helpful discussions.
GC acknowledges the warmfull hospitality of the OWL Institute at the University of Santa Cruz, where part of this work was accomplished.
This work was supported by the Programme National de Plan\'etologie (PNP) of
CNRS-INSU co-funded by CNES.

\bibliography{biblio_DC18}

\end{document}